\pdfoutput=1

\documentclass[12pt]{article}
\usepackage[
top = 2.5cm, 
bottom = 2.5cm,  
left = 2.55cm,  
right = 2.55cm]{geometry}

\usepackage{latexsym}  
\usepackage{verbatim}
\usepackage{tikz}
\usepackage{caption}
\usetikzlibrary{matrix}
 
\usepackage{color}
\usepackage{graphicx,amssymb,amsfonts,amsmath,amssymb,amscd,amstext, mathrsfs}
\usepackage{graphicx} 
\usepackage{bbm}
\usepackage{dsfont}
\usepackage{xcolor}
\usepackage[colorlinks=true, citecolor=black, linkcolor=black, allcolors=black]{hyperref}   
\usepackage{amsmath}
\usepackage{empheq}
\usepackage{hyperref}

\newlength\dlf

\def\be{\begin{eqnarray}}
\def\ee{\end{eqnarray}}
\def \bea {\begin{equation}} 
\def \eea {\end{equation}}

\newcommand{\CO}{{\cal O}}

\def \nn {\nonumber}

\def \rr {\raise.35ex\hbox{\small $\prime$}\kern-.17em{\mbox{\large $\imath$}}}
\def \del {\partial}
\def \dels {\partial\kern-.5em / \kern.5em}
\def \As {{A\kern-.5em / \kern.5em}}
\def \Ds {D\kern-.7em / \kern.5em}

\def\frac#1#2{{#1\over #2}}

\newcommand{\<}{\langle}
\renewcommand{\>}{\rangle}

\def \iffa {\iffalse} 

\def \ed  {\end{document}}

\def\nn{\nonumber}

\def \ed  {\end{document}}

\usepackage{cite}
\begin{document}
\thispagestyle{empty}
\begin{flushright} 
\end{flushright}

\begin{center} 

{\LARGE{\textsc{\bf{
Probing Universalities in $d>2$ CFTs:\\
\vspace{0.4cm} 
from Black Holes to Shockwaves
}}}}
\vspace{1.4cm}  

A. Liam Fitzpatrick$^*$,
Kuo-Wei Huang$^*$, 
and Daliang Li$^+$
\\ 
\vspace{1.2cm} 
{\it
$^*$Department of Physics, Boston University,\\
Commonwealth Avenue, Boston, MA 02215, USA\\
\vspace{0.5cm} 
$^+$Center for the Fundamental Laws of Nature,\\
 Harvard University, Cambridge, MA 02138, USA}
\end{center}
\vspace{2cm}

\noindent{Gravitational shockwaves are insensitive to higher-curvature corrections in the action.  Recent work found that the OPE coefficients of lowest-twist multi-stress-tensor operators, computed holographically in a planar black hole background, are insensitive as well.  In this paper, we analyze the relation between these two limits.  We explicitly evaluate the two-point function on a shockwave background to all orders in a large central charge expansion.  In the geodesic limit, we find that the ANEC exponentiates in the multi-stress-tensor sector.  To compare with the black hole limit, we obtain a recursion relation for the lowest-twist products of two stress tensors in a  \textit{spherical} black hole background, letting us efficiently compute their OPE coefficients and  prove their insensitivity to higher curvature terms.  After resumming  the lowest-twist stress-tensors and analytically continuing their contributions to the Regge limit, we find a perfect agreement with the shockwave computation.
 We also discuss the role of double-trace operators, global degenerate states, and multi-stress-tensor conformal blocks. 
These holographic results suggest the existence of a larger universal structure in higher-dimensional CFTs.

}

\newpage
\tableofcontents
\newpage

\addtolength{\parskip}{0.5 ex}
\jot=1.5 ex

\section{Introduction and Summary}

Conformal Field Theories (CFTs) have a relatively rigid structure compared to generic quantum field theories (QFTs), which can be exploited to study some of the behavior of QFT at strong coupling.  Our understanding of this structure and its consequences has advanced rapidly in recent years, especially in spacetime dimension $d>2$, and it is likely that there remains vastly more to learn.  One indication of our relative ignorance is that solutions to the constraints of crossing symmetry, as manifested in the bootstrap equation, emerge  from numerical analyses as points in parameter space that miraculously survive even as the regions surrounding them are shown to be inconsistent with basic CFT principles.  In rational 2d CFTs, such miracles are understood as consequences of shortening conditions of representations of the conformal algebra at special values in the space of conformal dimensions and central charge.  The modes of the stress tensor in 2d are the conformal generators and therefore central to such constructions.

In addition to studying  general CFTs, one may obtain stronger constraints by making ``sparseness'' conditions on the dimensions $\Delta$ of operators in the theory.  Such sparseness conditions are generally statements forbidden certain operators with dimensions below a ``gap'' dimension chosen by hand.  In the mildest cases, they might be simply that a certain Operator Product Expansion (OPE)  contains no relevant (i.e. $\Delta< d$) operators, whereas in the strongest cases they might forbid all but a few operators in the theory to have dimensions below a parametrically large gap.  These assumptions often lead to interesting consequences, and can drastically simplify the allowed space of CFTs.  An important source of intuition and computations about sparseness comes from the Anti de Sitter (AdS)/CFT correspondence, where gaps in the spectrum of operators translate into gaps in the masses of bulk fields.  Effective Field Theory (EFT) ideas applied to bulk theories elegantly predict and explain the relative simplicity of CFTs with very sparse spectra \cite{Heemskerk:2009pn, Fitzpatrick:2010zm} 
as simple consequences of the suppression of irrelevant interactions in the bulk. 
Some EFT type constraints are not obvious to derive from the CFT, but can be made manifest using unitarity and causality. 
For instance, the structures of the three-stress tensor coupling can deviate from the prediction of bulk Einstein gravity, 
but the deviation is perturbatively suppressed by the large gap $~\frac{1}{\Delta_{{\rm gap}}^\#}$. 
This was first shown  in \cite{Camanho:2014apa} using multiple 
small shockwaves in the AdS and further proved in the CFT \cite{Afkhami-Jeddi:2016ntf,Kulaxizi:2017ixa,Li:2017lmh,Afkhami-Jeddi:2017rmx}. 
Similar perturbative suppression also happens to the coupling between the 
stress-tensor sector and
other single-trace states in the CFT \cite{Meltzer:2017rtf,Afkhami-Jeddi:2018apj}.

The $d>2$ CFT structures we will be exploring go beyond the perturbative universality described above. 
They correspond to properties in AdS gravity that do not receive any perturbative $1/\Delta_{{\rm gap}}$ corrections. 
In this paper, we focus on two such properties: the near boundary behavior of the AdS-Schwarzschild solutions and large gravitational shockwaves. 
Recently, using the blackhole solutions, \cite{Fitzpatrick:2019zqz} found that higher-curvature terms do not affect the OPE coefficients 
of the ``lowest twist''\footnote{Recall that the twist of an operators is defined as its dimension minus its spin.} 
products of the CFT stress tensor.  
On the other hand, it was realized some time ago that higher-curvature terms 
do not affect the form of gravitational shockwave solutions in AdS \cite{Horowitz:1999gf}. 
These two results indicate that CFT four-point functions in two different limits 
are ``universal'' in the sense that they do not depend on 
effects that can be absorbed into higher-curvature corrections.\footnote{For related recent discussions, see \cite{Karlsson:2019qfi, Li:2019tpf, Kulaxizi:2019tkd}.}   
Any corrections to these CFT quantities 
must be suppressed non-perturbatively as $\sim e^{-\Delta_{{\rm gap}}}$, suggesting a greater robustness at large but finite $\Delta_{{\rm gap}}$. 
We will explore the relation between the universality of the black-hole 
and the shockwave, emphasizing where they overlap and where they are complementary.\footnote{The analogous connection between leading twist operators and Regge limits in the context of single-trace operators at weak coupling was analyzed in \cite{Costa:2013zra}.  }  
Optimistically, their existence hints at a larger, coherent structure  
which contains them both.

\subsection*{Summary}

Our main analysis is as follows.  In all cases, we study facets of the heavy-light 
correlator 
\be
\< \CO_L \CO_L \CO_H \CO_H\>
\ee of two light scalar operators $\CO_L$ 
with dimension $\Delta_L$ and two heavy scalar operators $\CO_H$ with dimension $\Delta_H$.
We use bulk gravity to compute the two-point function of $\CO_L$ in the background 
metric produced by the heavy operator, from which we 
extract universal pieces of the OPE data of the CFT.    

In the black hole regime, the ``heavy-light'' limit is defined as 
\be
{\rm Heavy{\text -}light~ limit}:~~~ C_T \rightarrow \infty ~~\textrm{with}~~\Delta_L,~  {\Delta_H\over C_T}~~ {\rm fixed}  \ .
\ee
The heavy operator $\CO_H$ then creates a black hole geometry that depends on $\Delta_H$ as well as all the higher-curvature terms in the gravity action. 
The two-point function of $\CO_L$ in this background can be interpreted as a heavy-light four-point function:\footnote{The $\CO_L$ two-point 
function in the black hole background is really a thermal average over the heavy 
operators in the four-point function.  For the composites of stress tensors 
that we consider at large $C_T$, the OPE coefficients can be determined 
solely from a near-boundary expansion of the bulk metric \cite{Fitzpatrick:2019zqz}, which should be insensitive to this distinction.}
\be
G(z,\bar{z}) \equiv \< \CO_L(1) \CO_L(1-z,1-\bar{z})\>_{\rm BH} 
\cong \< \CO_H(\infty) \CO_L(1) \CO_L(1-z,1-\bar{z}) \CO_H(0)\> \ .
\ee
We can extract OPE coefficients by performing a conformal block decomposition of this correlator:
\be
G(z,\bar{z}) = \sum_{\CO} c_\CO g_{\Delta_\CO, J_\CO}(z,\bar{z}) \ ,
\ee
where the sum over $\CO$ is a sum over primary operators, $g_{\Delta_\CO, J_\CO}(z,\bar{z})$ 
are conformal blocks,  and the coefficient $c_\CO = c_{LL \CO} c_{HH \CO}$ is the product of 
OPE coefficients of $\CO$ inside the $\CO_L \times \CO_L $ OPE and inside the $\CO_H \times \CO_H$ OPE. 
At leading order in the heavy-light limit, the operators  $\CO$ that contribute are ``double-trace'' operators 
made from two $\CO_L$s together with derivatives, as well as ``$T^n$'' operators made from $n$ products 
of the stress tensor.  The dimensions of the double-trace operators are $2\Delta_L$ plus integers, whereas 
the dimensions of the $T^n$ operators are integers, and therefore can be cleanly separated as 
long as $\Delta_L \notin \mathbb{Z}$. Moreover, at each $n$, the lowest possible twist of the $T^n$ operators 
is $\tau_{\rm min}(n) = n(d-2)$.  For each $n\ge 2$, there are an {\it infinite} number of such operators with spin 
increasing from $2n$ to $\infty$. 
 In all our calculations so far, we have found that the OPE coefficients of these 
``lowest-twist'' $T^n$s are universal in the sense that they are fixed by $\Delta_L, \Delta_H$ and $C_T$, 
independently of the higher-curvature terms in the gravity action.   
In particular, the ratio
\be
f_0 = \frac{4 \Gamma(d+2)}{(d-1)^2 \Gamma^2(\frac{d}{2})} \frac{\Delta_H}{C_T} \ ,
\ee 
(in units of the AdS radius of curvature) defined
as the coefficient of the first-order correction to the bulk metric in an expansion 
near the AdS boundary plays an important role for us. In $d=2$, the resummation of lowest-twist $T^n$ operators is the holomorphic part of the Virasoro vacuum block, which is completely determined by the Virasoro algebra and encapsulates many important aspects of quantum gravity in AdS$_3$  (e.g. \cite{Fitzpatrick:2016ive,Fitzpatrick:2015zha,Fitzpatrick:2014vua,Chen:2017yze,Hartman:2013mia,Roberts:2014ifa,Anous:2016kss}).  In $d>2$, the behavior of such operators, even in the heavy-light limit, is much less well understood.  We will focus on $d=4$ for concreteness but our methods should generalize to other dimensions. 

In this paper, we give a proof of this universality for $n=2$ with a {\it spherical} black hole.\footnote{It would be 
more satisfactory to extend this proof to general $n$, but we shall not do so in this work. 
The general proof with a planar black hole was given in \cite{Fitzpatrick:2019zqz}.}   
In this case, the lowest-twist $T^n$ dominate over those with higher twists in the limit  
\be
{\rm Lowest{\text-}twist~ limit}:~~~\bar{z} \rightarrow 0~~ {\rm with} ~~\frac{\Delta_H}{C_T} \bar{z} ~~{\rm fixed} \ .
\ee
Compared to \cite{Fitzpatrick:2019zqz}, which mostly considered taking 
$z$ small as well, we here keep $z$ arbitrary.  This limit has the advantage that it retains more information, and 
allows us to explore the behavior of the $T^n$ contributions in a wider regime.  
In particular, it allows us to make a connection to shockwaves.  

We find that the correponding lowest-twist $T^2$ OPE coefficients in $d=4$ 
 take the form
\be
c_{T^2, J} &=& \Delta  f_0^2 
\frac{J(J-2) \Gamma (J-3) \Gamma (J+1)}{(J+2) (J+4) (J+6) \Gamma (2 J+2)}
\frac{ 9 (\Delta +1) (\Delta +2)+\Delta  (\Delta +1) x_J+\Delta  (\Delta +2) y_J}{\Delta -2} \nn\\
 x_J &=& \frac{3}{16} (J-1) (J-3) (J+4) (J+6) \ , ~~~~ y_J = -\frac{1}{8} (J-2) (J-3) (J+5) (J+6),
 \label{eq:T2OPE}
\ee
for $J=4,6,8, \dots$ ($\Delta \equiv \Delta_L$ for compactness).
As $z$ is arbitrary, it can be analytically continued through a branch cut passing from $z=1$ to $\infty$ 
to the second sheet of the correlator. We can then study the ``Regge'' limit, where 
\be
{\rm  Regge~limit}:~~ z \in {\rm 2nd \text{-} sheet}\ , ~~z,\bar{z} \to 0 ~~{\rm{with }}~ {z\over \bar{z}}~ {\rm{fixed}} \ .
\ee

Our present black-hole computation lets us resum only the lowest-twist $T^2$s and thus it probes the 
Regge limit only at $\bar{z} \ll z \ll 1$. 
Moreover, the black-hole approach does not determine double-trace contributions as they require a horizon boundary condition.  
However, the full contribution in the Regge limit for general $\bar{z}/z$, including double-traces, can    
be computed using instead a shockwave background.   
Based on an earlier work \cite{Cornalba:2006xk,Cornalba:2007zb,Cornalba:2006xm,Cornalba:2008qf}, which extended Eikonal methods to AdS/CFT correlators, 
 we will 
explicitly evaluate $G(z,\bar{z})$ in the Regge limit to {\it any~order} in $\frac{1}{C_T}$;  
the result in $d=4$ is
\be
&&G_{\rm Eik}(z,\bar{z}) = \sum_{n=0}^\infty   \frac{(3 \pi i f_0)^n}{(z \bar{z})^{\Delta+\frac{n}{2}} }
\frac{(-1)^n \Gamma (1-n) \Gamma \left(2 \Delta+n\right)   \Gamma \left(2 n+\Delta-1\right)}{\Gamma (n+1) \Gamma \left(\Delta-1\right){}^2
   \Gamma \left(\Delta+n\right) \left(2 \Delta+n-1\right)}
   \label{eq:EikHnIntro}\\
&& \times {\eta ^{\Delta+\frac{n}{2}}\over {\eta-1}}  \Big(F_{\Delta,n,-1}(\eta )+\frac{(\eta -1) \left(\Delta+2 n-1\right) F_{\Delta ,n,0}(\eta )}{\Delta+n}\nn\\
&&~~~~~~~~~~~~~~~~~~~~~~~~~~~~ -\frac{\eta 
   \left(\Delta+2 n-1\right) \left(\Delta+2 n\right) F_{\Delta,n,1}(\eta )}{\left(\Delta+n\right) \left(\Delta+n+1\right)}\Big) + \left( \eta \rightarrow \eta^{-1 } \right) \nn,
\label{eq:ShockwaveAllOrders}
\ee
where $\eta \equiv \frac{\bar{z}}{z}$
and $F_{\Delta, n,a}$ is a hypergeometric function defined in (\ref{eq:SWHyp}). 
To compare with our black-hole background method, we will extract the $T^n$ 
contributions and take the small $\bar{z}/z$ limit, with the result
 \be
G_{\rm Eik}(z,\bar{z})|_{C_T^{-n}} ~\propto~ \left( f_0 \frac{\bar{z}}{z^2} \right)^n    
\frac{  \Gamma \left(\Delta-n\right) \Gamma \left(2 n+\Delta-1\right)}{n! \Gamma \left(\Delta-1\right) \Gamma \left(\Delta\right)}\left( 1 + \CO\big(\frac{\bar{z}}{z}\big) \right) .
   \ee
Taking $n=2$, we find that the result exactly matches the resummation of $T^2$ operators from the black-hole method.
Moreover, we recognize the matching piece in section \ref{sec:ReggePole} as the contribution of a spin 3 null line operator on the second sheet. 
From the perspective of null line operators, the $O_HO_H$ OPE and the $O_L O_L$ OPE each contributes an ANEC. The OPE of these two ANEC operators contain a spin-3 operator \cite{Hofman:2008ar,Kologlu:2019mfz} whose contribution is observed here.

Aside from its relation with the black-hole method analysis, the all-orders 
result (\ref{eq:EikHnIntro}) is interesting in its own right.  As we discuss in section \ref{sec3}, in the geodesic 
limit $C_T, \Delta_L , \Delta_H \rightarrow \infty, z , \bar{z} \rightarrow 0$ 
with $\bar{z}\over z$ and $\frac{\Delta_L \Delta_H}{C_T z}$ fixed, the result takes the form
\be
\ln G_{\rm Eik} 
~\propto~ - \frac{\Delta_H \Delta_L \bar{z}}{C_T (z-\bar{z})z}  
~\propto~ \left\<\CO_H  \left( \int_{-\infty}^\infty dx^- T_{--} \right)  \CO_H\right\>.
\label{eq:LeadingSing}
\ee
The last term above is the averaged null energy (ANEC) operator in the heavy-state background.  
We may interpret the behavior of $G_{\rm Eik}$ in this limit as the exponentiation of the ANEC.   

Another interesting feature of (\ref{eq:EikHnIntro}) is the pattern of zeros and poles in the coefficients 
of the leading singularity (\ref{eq:LeadingSing}).  The zeros are very similar to those that appear in 
$d=2$ \cite{Chen:2016cms}, and we make some comments about how a simple derivation 
of these factors in $d=2$ might generalize to higher dimensions. 

\subsection*{Outline}

We first discuss basic relations between a shockwave and a black hole 
in Sec. \ref{sec2}.  A more detailed analysis on shockwaves and the explicit 
Eikonal resummation are presented in Sec. \ref{sec3}. 
In this section, we also make a couple of remarks for $d>2$ CFTs. 
In Sec. \ref{sec4}, we prove the universality of the lowest-twist operators 
with double stress-tensors with a spherical black hole, focusing on $d=4$.  
We will provide a closed form for the corresponding lowest-twist OPE coefficients. 
After analytically continuing to the second sheet, we resum the contributions 
and compare with the shockwave computation. 
We conclude with some future directions.  
An appendix includes some technical details related to Sec. \ref{sec4} 
and several lowest-twist OPE coefficients.

{\it Note: while this paper was in preparation, \cite{Kulaxizi:2019tkd} appeared on arXiv, which 
also contains the formula (\ref{eq:T2OPE}) and discusses the matching between the 
Eikonal limit and the analytic continuation of the black-hole limit at $\CO(\frac{1}{C_T^2})$.}

\section{Shockwaves and Black Holes}
\label{sec2}

A gravitational shockwave is created when a massive source is boosted
close to the speed of light. This includes, for example, boosting
a black hole. A large boost increases the gravitational backreaction
of the source, so naively taking the large boost limit might not result
in a well-defined geometry. In a limit where the rest mass of source
scales inversely  with the boost factor, the resulting goemetry may
still be singular, but it is geodesic complete and all its geodesics
will only be shifted a finite distance away from the vacuum geodesics. 
We shall refer to this as the shockwave limit, and the
resulting geometry the shockwave solution.

The shockwave solutions in  flat space were discovered in \cite{Aichelburg1971}. 
They were generalized to AdS$_{4}$ in \cite{Hotta_1993}, and 
to general dimensions \cite{Sfetsos:1994xa, Horowitz:1999gf}.  
A salient nature of shockwave solutions is that they are universal in that they are insensitive to
higher-order derivatives and/or stringy corrections to gravity \cite{Horowitz:1999gf, PhysRevLett.64.260, AMATI1989443}. 
Extracting what this shockwave universality implies for the CFT data and how 
is it related to the universality in the black-hole background is the main motivations of the present work.    

As a warm-up, we briefly review the properties of shockwave solutions and
their relations to black hole solutions in this section. 
We first introduce the embedding space for AdS$_{d+1}$ and show how to create a shockwave
by boosting a small black hole, following \cite{Hotta_1993}. 
We next discuss the universality of the shockwave and make preliminary comments 
on its relation to the universality of black holes.  In particular, we shall see that the leading 
order in the large boost limit for a shockwave is related 
to the leading order in the large $r$ limit for a black hole. 
We leave more detailed discussions to later sections.   

\subsection{Embedding Space}

It is convenient to describe the Lorenzian   
AdS$_{d+1}$ space in terms
of an $d+2$ dimensional embedding space $\mathbb{M}_{2}\times\mathbb{M}_{d}$
with signature $(2,d)$,
\begin{equation}
ds^{2}=-dx_{+}dx_{-}-dxd\bar{x}+\sum_{i=1}^{d-2}dx_{i}^{2} \ .
\label{eq:embedding LC}
\end{equation}
We denote vectors in this embedding space using capital
letters. AdS$_{d+1}$ is represented by 
$X=\left(x_{+},x_{-},x,\bar{x},x_{1},\dots x_{d-2}\right)\in\mathbb{M}_{2}\times\mathbb{M}_{d}$
satisfying $X^{2}=-1$; the boundary of AdS$_{d+1}$ is represented
as $P^{2}=0$ with $P\sim\lambda P$ identified, where $\lambda$
is any real positive number. 
In the embedding space, the AdS$_{d+1}$
isometry and the corresponding conformal transformation in the CFT
can be realized as the same linear transform in $\mathbb{M}_{2}\times\mathbb{M}_{d}$. 

The empty AdS$_{d+1}$ metric 
is
\begin{equation}
ds_{0}^{2}=-\left(r^{2}+1\right)dt^{2}+\frac{dr^{2}}{r^{2}+1}+r^{2}d\Omega_{d-1},
\end{equation}
where $d\Omega_{d-1}$ is the volume form of the $d-1$ dimensional
unit sphere. One can relate the embedding space coordiante to the 
spherical 
coordinate via the map
\begin{equation}
X=\left(x_{-1},x_{0},x_{1},\dots,x_{d}\right)=\left(\sqrt{1+r^{2}}\cos t,\sqrt{1+r^{2}}\sin t,r\hat{x}\right) \ ,
\end{equation}
where $\hat{x}\in\Omega_{d-1}$ is a $d$-dimensional unit vector.
This transforms the empty AdS metric to 
\begin{equation}
ds_{0}^{2}=-dx_{-1}^{2}-dx_{0}^{2}+\sum_{i=1}^{d}dx_{i}^{2}  \ .
\end{equation}
 It is easy to check that $X^{2}=-1$ using this metric. It is related
to the lightcone coordinate used in (\ref{eq:embedding LC}) by
$x_{\pm}=x_{0}\pm x_{d-1}$, $x=x_{-1}+x_{d}$ and $\bar{x}=x_{-1}-x_{d}$. 

Relatedly, another useful coordinate system is the Poincare coordinate,
in which the empty AdS metric takes the form
\begin{equation}
ds_{0}^{2}=\frac{1}{y^{2}}\left(dy^{2}-dudv+{\displaystyle \sum_{i=1}^{d-2}}dw_{i}^{2}\right)  \ ,
\end{equation}
where $y\in\left(0,\infty\right)$ is the radial coordinate with $y=0$
being the boundary. Here 
\begin{equation}
\left(x_{+},x_{-},x,\bar{x},x_{1},\dots , x_{d-2}\right)
=\frac{1}{y}\left(y^{2}-uv+{\displaystyle \sum_{i=1}^{d-2}}w_{i}^{2},1,u,v,w_{1},\dots , w_{d-2}\right) \ .
\end{equation}
A boost in AdS along a boundary direction can be represented as
$\left(u,v\right)\rightarrow\left(\lambda u,\lambda^{-1}v\right)$
with other coordinates fixed. This is also a boost in the embedding
space with $\left(x,\bar{x}\right)\rightarrow\left(\lambda x,\lambda^{-1}\bar{x}\right)$.
Note that the embedding space is symmetric under the exchange
of $\left(x,\bar{x}\right)\leftrightarrow\left(x_{+},x_{-}\right)$
while there is no such a manifest invariance in the Poincare coordinate.

\subsection{Creating a Shockwave by Boosting a Small Black Hole}

We here demonstrate that a shockwave metric can be created
by boosting a small black hole.  See \cite{Hotta_1993} for more details. 
We focus on $d=4$ but the method generalizes. 

The AdS$_{5}$-Schwarzschild metric reads
\begin{equation}
ds^{2}=-\left(r^{2}+1-\frac{2M}{r^2}\right)dt^{2}+\frac{1}{r^{2}+1-\frac{2M}{r^2}}dr^{2}+r^{2}d\Omega_{3}^{2} \ .
\end{equation}
As explained above, we shall take the small $M$ limit, where
\begin{equation}
ds^{2}=ds_{0}^{2}+Mds_{1}^{2}+\mathcal{O}\left(M^{2}\right) 
\label{eq:metricPerturbation}\ , ~~~ 
ds_{1}^{2}=\frac{2}{r^2}\left(dt^{2}+\frac{1}{\left(1+r^{2}\right)^{2}}dr^{2}\right) \ .
\end{equation} 
A boost $\left(x,\bar{x}\right)\rightarrow\left(\lambda x,\lambda^{-1}\bar{x}\right)$
leaves $ds_{0}^{2}$ invariant and transforms $ds_{1}^{2}$. The
shockwave solution, which contains a Dirac $\delta$-function, will
emerge from the large boost limit of $\lambda\rightarrow0$. 
To understand
the appearance of this $\delta$-function,  consider a generic function   
$\Omega\left(x,\lambda\right)$ satisfying 
\begin{equation}
\Omega\left(x\neq0,0\right)=0,\hspace{1em} 
{\rm and} 
\hspace{1em}\Omega\left(\lambda\tilde{x},\lambda\right)=\frac{1}{\lambda}g\left(\tilde{x}\right)\ . \label{eq:deltaConditions}
\end{equation}
Thus,
\begin{equation}
\lim_{\lambda\rightarrow0}\Omega\left(x,\lambda\right)=\delta\left(x\right)\int_{-\infty}^{\infty}g(\tilde{x})d\tilde{x} \ .
\end{equation}
The boosted metric $ds_{1,b}^{2}$ takes the following form after
a further coordinate transformation with $x\rightarrow x-\lambda^{2}\bar{x}$:
\begin{equation}
ds_{1,b}^{2}=\frac{1}{\lambda}\left(\Omega_{1}dx^{2}+\lambda \Omega_{2}dxd\bar{x}+\lambda^{2}\Omega_{3}d\bar{x}^{2}\right)
\end{equation}
where $\Omega_{1,2,3}$ are functions satisfying (\ref{eq:deltaConditions}).
For instance,
\begin{equation}
\Omega_{1}=\frac{32m\left(\frac{\left(x_{0}^{2}+1\right)x^{2}}{\lambda^{2}}+4\left(x_{0}-1\right)x_{0}^{2}\left(x_{0}+1\right)\right)}{\lambda\left(\frac{x^{2}}{\lambda^{2}}+4x_{0}^{2}-4\right){}^{2}\left(\frac{x^{2}}{\lambda^{2}}+4x_{0}^{2}\right){}^{2}} \ ,~~~~ m= {M\over \lambda} \ .
\end{equation} 
The expressions of $\Omega_2$, $\Omega_3$ become lengthy and we will not list them here.  
In the large boost, small mass limit where $\lambda\rightarrow0$,
$m=\frac{M}{\lambda}$ fixed, the $dx^{2}$ term is the only surviving
piece. We obtain
\begin{equation}
\lim_{\lambda\rightarrow0,~M=\lambda m}ds_{1,b}^{2}=-\frac{3\pi m\left(2x_{0}\left(\sqrt{x_{0}^{2}-1}-x_{0}\right)+1\right)}{2\sqrt{x_{0}^{2}-1}}\delta\left(x\right)dx^{2}
\label{eq:ShockwaveMetricFromBoosting}
\end{equation}
With an infinite boost, a point mass generates a singular metric that remains well-behaved as it only causes a finite shift in the geodesic. 
The prefactor in (\ref{eq:ShockwaveMetricFromBoosting}) corresponds to a propagator in the transverse hyperbolic space with $x=0$, which will be given later in (\ref{eq:hFunc}).

More generally,  one can consider 
\begin{equation}
ds^{2}=- \big(1+r^2 f\left(r\right)\big)dt^{2}+\frac{1}{1+r^2 h\left(r\right)}dr^{2}+r^{2}d\Omega_{3}^{2}
\end{equation}
as the spherically symmetric vacuum solution of a more general gravitational 
theory and adopt a near-boundary expansion\footnote{One can show by performing a conformal block decomposition that $h_0=f_0$ \cite{Fitzpatrick:2019zqz} if the stress tensor is the only dimension $ \le 4$ operator in the theory.}
\begin{equation}
f(r)=1-{f_0\over r^4}+\dots \ , ~~~h(r)=1-{h_0\over r^4}+\dots
\end{equation}
with possible higher-order corrections. 
However, as the metric $ds_{1}^{2}$ defined in (\ref{eq:metricPerturbation}) is not affected by higher-order 
corrections, the shockwave solution is manifestly universal.

\section{Eikonal Limit Resummation}
\label{sec3}

In this section, starting with the results of \cite{Cornalba:2006xk}, we will 
obtain the Eikonal limit of $d=4$ heavy-light four-point functions from AdS$_5$, to all 
orders in $\frac{1}{C_T}$. 
We will compare to the computation of the lowest-twist stress tensors from a black hole background in Euclidean space at $\frac{1}{C_T^2}$.  
At this order $\frac{1}{C_T^2}$, only $T^2$ operators contribute, but already 
that will be enough to see explicitly how the lowest-twist black hole and shockwave computations overlap.  
The generalization to any higher order should be 
straightforward in principle, though computationally more involved.  

\subsection{Eikonal AdS/CFT} 

We begin by briefly reviewing the relevant results of \cite{Cornalba:2006xk}; see there for more details.  
The main physical point is that the Eikonal approximation for CFT heavy-light four-point 
correlators can be computed by treating the two heavy operators in the correlator as 
sources of a shockwave, and solving for the two-point function for the remaining two light 
operators in the shockwave geometry (see Fig-1).  
The two heavy operators are almost null 
separated from each other and generates a shockwave on a null sheet.

\begin{center}

\begin{tikzpicture}

\draw  (-2,2) -- (-2,0) arc (180:360:2cm and 0.5cm); -- (2,2) ++ (-2,-1) circle (2cm and 0.5cm);
\draw [rotate= -45] [fill=black!1!red, opacity=0.4] (-3.4 , -0.6, -5.2) ellipse (12pt and 80pt);
\draw [densely dashed] (-2,0) arc (180:0:2cm and 0.5cm);
\draw (-2,4) -- (-2,0) arc (180:360:2cm and 0.5cm) -- (2,4) ++ (-2,0) circle (2cm and 0.5cm);

\draw [draw=black!10!red,dashed, thick]  (2,4) node{~~~~~~~~$\color{red}{\rm P}_{{\rm H}_2}$} -- (-2,0) node{$\color{red}{\rm P}_{{\rm H}_1}$~~~~~~~~};
\draw [black!10!red,->,>=stealth] (1, 3) -- (1.3, 3.3);
\draw [black!10!red,->,>=stealth] (-1.3, 0.7) -- (-1, 1);

\draw [black!50!green, dotted, thick] (2,0) node{~~~~~~~~${\rm P}_{{\rm L}_1}$}  to  [bend angle = 8.1, bend right] (-2,4) node{${\rm P}_{{\rm L}_2}$~~~~~~~~};
\draw [black!50!green, ->,>=stealth]  (-1.2, 3.37) -- (-1.3, 3.45);
\draw [black!50!green, ->,>=stealth]  (1.4, 0.77) -- (1.3, 0.9);

\node at (-2,0) [circle, draw=black!10!red, fill=black!10!red, scale=0.25] {};
\node at (2,4) [circle, draw=black!10!red, fill=black!10!red, scale=0.25] {}; 
\node at (2,0) [circle,draw=black!50!green, fill=black!50!green,scale=0.25] {};
\node at (-2,4) [circle,draw=black!50!green, fill=black!50!green,scale=0.25] {};

\draw [orange, very thick]  (0, 1.4) -- (0, 2.55);

\usetikzlibrary{shapes.misc}
\tikzset{cross/.style={cross out, draw=black!10!red, minimum size=2*(#1-\pgflinewidth), inner sep=0pt, outer sep=0pt},cross/.default={2pt}}
\draw (0,2) node[cross,rotate=20] {};
\tikzset{cross/.style={cross out, draw=black!60!green, minimum size=2*(#1-\pgflinewidth), inner sep=0pt, outer sep=0pt},cross/.default={2pt}}
\draw (0,2.33) node[cross,rotate=20] {};

\end{tikzpicture}

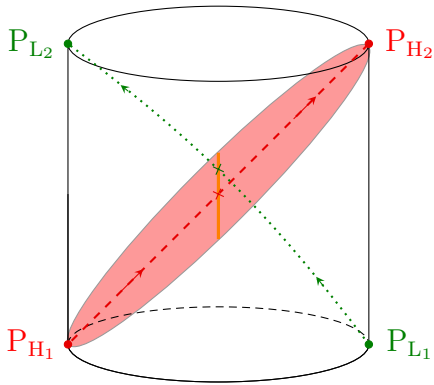
\captionof{figure}{\text{Operators in the shockwave geometry   
}}

\end{center}

 In  embedding space notation, we can take the two light operators to be at $P_{{\rm L}_1}, P_{{\rm L}_2}$ with 
   \be
 P_{{\rm L}_1} = (1,0,0,0,0,0), \qquad P_{{\rm L}_2} = (q \bar{q}, -1,  \mathbf{q}), ~~~~~ \quad \mathbf{q} = -(q,\bar{q},0,0) \ ,
 \ee
 and the two heavy source operators to be at $P_{{\rm H}_1}, P_{{\rm H}_2}$ with
 \be
 P_{{\rm H}_1} = (0,1,0,0,0,0), \qquad P_{{\rm H}_2} = (-1, p\bar{p}, \mathbf{p}), ~~~~~ \quad \mathbf{p} = -(p,\bar{p},0,0) \ .
 \ee
To connect to standard representations of the four-point function in terms of the $d=4$ 
positions of the operators, we need the following formulae for the $z,\bar{z}$ coordinates in terms of $\mathbf{p}, \mathbf{q}$:
\be
 z \bar{z} = \mathbf{q}^2 \mathbf{p}^2 = p \bar{p} q \bar{q} \ ,~~~~ z+\bar{z} = 2 \mathbf{p} \cdot \mathbf{q} = p \bar{q} + q \bar{p} \ .
\ee
Without loss of generality, we take $p= \bar{p}=-1$ to get the simple relations 
\be
q = z, ~~~~~ \bar{q} = \bar{z} \ .
\ee   

The light operator two-point function in the shockwave background follows 
from matching the standard AdS bulk-to-boundary propagator across the shockwave itself.  
As a result, the AdS computation of the Eikonal heavy-light correlator $G_{\rm Eik}(z,\bar{z})$ 
 reduces to an integral over the shockwave surface:  
\be
\label{Joao}
G_{\rm Eik}(z,\bar{z}) = \frac{\Gamma(2 \Delta)}{\Gamma^2(\Delta-1)} \int_{H_3} 
 \frac{d^3 \mathbf{x}}{(-2 \mathbf{q} \cdot \mathbf{x} + h(\mathbf{x} \cdot \mathbf{p}) + i \epsilon)^{2 \Delta}}  \ .
\ee
The coordinate $\mathbf{x}$ parameterizes the shockwave surface, which is described by 3d hyperbolic space $H_3$:
\be
\mathbf{x}= -(x,\bar{x},x_1, x_2), \quad \mathbf{x}^2 
= -1, \qquad \int_{H_3} d^3 \mathbf{x} 
= \int dx d\bar{x} dx_1 dx_2 \delta(\mathbf{x}^2+1) \ .
\ee 
The function $h$ is the metric deformation in the shockwave background.  In the limit 
where the shockwave is created by heavy operators, it is given by the propagator 
in hyperbolic space:\footnote{Although we are focusing on $d=4$, at this point very little 
changes for general $d$.  Aside from the change in the number of components of vectors, the 
only difference in general $d$ is the prefactor in $G_{\rm Eik}$, and the propagator in hyperbolic space:
\be
G_{\rm Eik}(z,\bar{z)} &=& \frac{\Gamma(2 \Delta)}{\Gamma^2(\Delta+1 - \frac{d}{2})} \int_{H_{d-1}} 
\frac{d^{d-1} \mathbf{x}}{(2 \mathbf{q} \cdot \mathbf{x} + h(\mathbf{x} \cdot \mathbf{p}) + i \epsilon)^{2 \Delta}}, \nn\\
 h(\mathbf{x} \cdot \mathbf{p}) &=& A_d  z^{-(d-1)} {}_2F_1\left(d-1,\frac{d+1}{2}, d+1,\frac{1}{z}\right) \ .
\ee 
}
\be
h(\mathbf{x} \cdot \mathbf{p}) = A  z^{-3} {}_2F_1\left(3,\frac{5}{2}, 5,\frac{1}{z}\right), 
\label{eq:hFunc}
\ee
 where $A\equiv A_4=\frac{-5 i \pi }{2} \frac{\Delta_H}{C_T}$,  
 and $z$ is the chordal distance between $\mathbf{x}$ and $\mathbf{p}$:
\be
 z \equiv -\frac{(\mathbf{x}-\mathbf{p})^2}{4} = \frac{1+\mathbf{x} \cdot \mathbf{p}}{2} 
= \frac{1}{2} \left( 1+ \frac{x + \bar{x}}{2} \right).
\ee

\subsection{Integrating Over the Shockwave Surface}

Our task now is to explicitly evaluate the integral 
\eqref{Joao} to all orders in an expansion in powers of $h$.   
As the only dependence in the integral on the coordinates $x_1, x_2$ is through the 
$\delta$-function, we can eliminate them immediately:
\be
\int dx_1 dx_2 \delta(-x \bar{x} +x_1^2 + x_2^2 +1) = \pi \Theta(x \bar{x}-1) \ .
\ee
We still have to integrate over $x$ and $\bar{x}$.  
A convenient choice of coordinates is
\be
x  + \bar{x} = 2 \cosh \chi, \qquad x \bar{x} = \cosh^2 t \ . 
\ee
This set of coordinates has the advantage that the condition $x \bar{x} \ge 1$ is 
automatically satisfied for $t$, and furthermore the propagator reduces to 
\be
h(\mathbf{x} \cdot \mathbf{p}) = 64 A \frac{e^{-|\chi|}}{e^{2|\chi|}-1} \ .
\ee
Note that the inverse change of coordinates has two branches
\be
x , \bar{x}  = \cosh \chi \pm \sqrt{\cosh^2 \chi-\cosh^2 t} \ ,
\ee
and each branch covers only half of the original space.  To cover the full space, we 
have to include both branches, which are related by $x \leftrightarrow \bar{x}$.  
Because the integrand is invariant if we perform $x\leftrightarrow \bar{x}$ together 
with $z \leftrightarrow \bar{z}$, we can integrate over only a single branch and then 
symmetrize on $z \leftrightarrow \bar{z}$. It will be convenient to define the ratio
\be
\eta \equiv \frac{\bar{z}}{z},
\ee
and $z \leftrightarrow \bar{z}$ symmetrization means $\eta \leftrightarrow \eta^{-1}$.  

Changing variables in the integral and expanding in powers of $h$, we find, at a given $n$,    
\be
&& G_{\rm Eik}(z,\bar{z})|_{h^n} = \frac{1}{(z \bar{z})^{\Delta+\frac{n}{2}} }
\frac{(-1)^n \Gamma \left(n+2 \Delta \right)}{\Gamma (n+1) \Gamma^2 \left(\Delta -1\right)} \\
    && \times \int_0^\infty d \chi \int_{0}^\chi dt \frac{\sinh 2t \sinh \chi 
 \left( 64 A\frac{e^{-|\chi|}}{e^{2|\chi|}-1}\right)^n}{\sqrt{\cosh^2 \chi -\cosh^2 t} 
\left(2\cosh\left( \frac{\ln \eta}{2} \right) \sqrt{\cosh^2 \chi -\cosh^2 t} +2 \cosh \chi\sinh \left( \frac{\ln \eta}{2} \right) \right)^{2\Delta+n} }\nn \\
    && + (\eta \leftrightarrow \eta^{-1}) . \nn 
\ee
Integrating over $t$ and $\chi$ gives 
\be
\label{eq:EikHn}
&&G_{\rm Eik}(z,\bar{z})|_{h^n} = \sum_{n=0}^\infty   \frac{(64 A)^n}{(z \bar{z})^{\Delta+\frac{n}{2}} }
\frac{(-1)^n \Gamma (1-n) \Gamma \left(2 \Delta+n\right)   \Gamma \left(2 n+\Delta-1\right)}{\Gamma (n+1) \Gamma \left(\Delta-1\right){}^2
   \Gamma \left(\Delta+n\right) \left(2 \Delta+n-1\right)}
   \\
&& \times {\eta ^{\Delta+\frac{n}{2}}\over {\eta-1}}  \Big(F_{\Delta,n,-1}(\eta )+\frac{(\eta -1) \left(\Delta+2 n-1\right) F_{\Delta ,n,0}(\eta )}{\Delta+n}\nn\\
&&~~~~~~~~~~~~~~~~~~~~~~~~~~~~ -\frac{\eta 
   \left(\Delta+2 n-1\right) \left(\Delta+2 n\right) F_{\Delta,n,1}(\eta )}{\left(\Delta+n\right) \left(\Delta+n+1\right)}\Big) + \Big( \eta \rightarrow \eta^{-1 } \Big) \nn,
\ee
where 
\be
F_{\Delta, n,a}(\eta) \equiv \, _2F_1\left(a+2 n+\Delta,n+2 \Delta;a+n+\Delta+1;-\eta \right).
\label{eq:SWHyp}
\ee

The result \eqref{eq:EikHn} is singular on the physical values 
$n \in \mathbb{N}$ due to the $\Gamma(1-n)$ factor.
This corresponds to the singularity in the propagator at $\chi \sim 0$, which is 
a UV divergence.  We shall regulate this divergence by analytic continuation in $n$, i.e. by 
subtracting off the pole at integer values of $n$ and keeping the finite piece.  However, this 
regularization is not necessary, as we will see,  if we restrict our attention to the 
contribution from $T^2$ operators: the singularity arises only in the OPE coefficients of the ``$\CO_L^2$'' double-trace  operators.

\subsection{Separating Double-Traces and $T^n$s}

The expression (\ref{eq:EikHn}) contains contributions from the exchange of 
double-trace operators made from $\CO_L$ as well as from stress tensors.  
One can separate out these two types of contributions in a small $\eta$ expansion 
based on whether the powers of $\eta$ are integers or $\Delta_L$ plus integers.\footnote{This 
distinction breaks down when $\Delta$ is itself an integer. 
}  

We see that all the stress-tensor contributions come from the 
$\eta \leftrightarrow \eta^{-1}$ piece in the last line of (\ref{eq:EikHn}), 
as the series expansion of the other pieces in (\ref{eq:EikHn}) manifestly contains only powers of 
$\eta$ of the form $\Delta +m$ for $m$ a certain integer because of the prefactor $\eta^{\Delta+\frac{n}{2}}$ 
and the fact that the hypergeometric functions have regular series expansions.

Let us convert the result into a form where the 
$T^n$ pieces can be read off easily. 
Use the following identity:  
\be
{}_2F_1(a,b,c,-z) &=& \frac{\Gamma(b-c)\Gamma(c)}{\Gamma(c-a)\Gamma(b)} z^{-a} {}_2F_1(a,a-c+1,a-b+1,-z^{-1}) \nn\\
&& + \frac{\Gamma(a-b)\Gamma(c)}{\Gamma(c-b) \Gamma(a)} z^{-b} {}_2F_1(b,b-c+1,b-a+1,-z^{-1}) \ .
\ee
The stress-tensor pieces can be extracted by taking the terms without $\eta^{\Delta}$ factors.  
We are most interested in the leading-twist contribution at each order in $h^n$: 
\be
\frac{G_{\rm Eik}(z,\bar{z})|_{h^n}}{G_{\rm Eik}(z,\bar{z})|_{h^0}} = 
\left( 3 i \pi f_0\right)^n 
\left( \frac{\bar{z}}{z^2} \right)^n \frac{  \Gamma \left(\Delta-n\right) 
\Gamma \left(2 n+\Delta-1\right)}{n! \Gamma \left(\Delta-1\right) \Gamma \left(\Delta \right)}\left( 1 + \CO\big(\frac{\bar{z}}{z}\big) \right) .
   \label{eq:LTfromSW}
   \ee
We have divided by the $n=0$ term, since this term corresponds to the identity 
exchange and therefore just sets the overall normalization of the external operators. 

The expression \eqref{eq:LTfromSW} is the leading contribution of the 
lowest-twist $T^n$ operators in the Regge limit with $\bar{z} \ll z$.  
In the next section, we will see how to reproduce the $n=2$ term by 
resummation and analytic continuation of an all-orders (in spin)  
computation of the lowest-twist $T^2$ operators in a black hole 
background, using the methods of \cite{Fitzpatrick:2019zqz}.

\subsection{Cancellation of Poles}

Although we focus in this work on the contributions from $T^n$ 
operators rather than from double-traces, they are clearly connected 
and it is interesting to understand when contributions from one can be 
used to determine contributions from the other.  One relatively simple 
connection between them is that the poles $\sim \frac{1}{\Delta -m}$ in 
the $T^n$ operator contributions at positive integer values $m$ must also be 
present with the opposite residue in the double-trace contributions.  
The reason for this relation is simply that the full correlator should not have 
such poles and therefore they must cancel between the two types of contributions.  
This relation was predicted in \cite{Fitzpatrick:2019zqz}, but it was difficult to check explicitly 
in that context, because in a black hole background the double-trace operator 
coefficients depend on a boundary condition at the black hole horizon.  
In the shockwave case, however, we have both the $T^n$ contributions and 
the double-trace contributions.  In principle, the necessary cancelation is already 
manifest in the expression for the four-point function by virtue of the fact that it 
does not have any poles at $\Delta=2,3,4, \dots$, so any poles in the $T^n$ piece 
obviously must cancel in the full correlator.  Let us here see this cancellation explicitly.

We can expand (\ref{eq:EikHn}) at small $\eta$ and observe the following two terms:
\be
 (z \bar{z})^{\Delta } G_{\rm Eik}\supset  \left( \frac{  -2^n  \sin ^2\left(\pi  \Delta \right)\Gamma \left(1-\Delta
   \right) }{n! \pi^2}\right) 
    \Big[ \eta ^{\Delta } \Gamma (2-n)  \Gamma \left(n-\Delta \right) \Gamma
   \left(n+2 \Delta -1\right)  \quad && \
 \nn\\
   + \eta^n  \Gamma \left(2-\Delta \right) \Gamma \left(\Delta -n\right) \Gamma
   \left(\Delta+2n -1\right) \Big]  \ . && 
   \ee
   The powers of $\eta$ identify the contribution in the first line as 
coming from a double-trace, and in the second line as coming from a $T^n$.  
At $n=2$, the double-trace contribution has a pole $\sim \frac{1}{n-2}$, which 
we discard as it is due to a UV divergence as mentioned earlier.  
Keeping the finite piece at $n=2$, the result has a pole at $\Delta = 2$ of the form
   \be
    (z \bar{z})^{\Delta } G_{\rm Eik}(z,\bar{z}) &\supset& - \frac{48 \eta^2}{\Delta-2} \qquad \textrm{(double-trace)} \ .
    \ee
    In addition, the second line has a pole at $\Delta=2$ of the form
 \be
    (z \bar{z})^{\Delta } G_{\rm Eik}(z,\bar{z}) &\supset&  \frac{48 \eta^2}{\Delta-2} \qquad  (T^2)\ ,
    \ee
    so the two poles can be seen explicitly to cancel, as required.

\subsection{Zeros and Degenerate States}

Observe that the $\Gamma$ function factors for the first few terms of (\ref{eq:LTfromSW}) are 
\be
n=0 :&& 1 \ , \nn\\
n=1 :&& \Delta \ , \nn\\
n=2 :&& \frac{\Delta \left(\Delta+1\right) \left(\Delta+2\right)}{2 \left(\Delta-2\right)}\ , \nn\\
n=3 :&&\frac{\Delta \left(\Delta+1\right) \left(\Delta+2\right) \left(\Delta+3\right) \left(\Delta+4\right)}{6 \left(\Delta-3\right)
   \left(\Delta-2\right)}\ .
\ee 
Clearly, the entire dependence of these factors on $\Delta\equiv \Delta_L$ is 
fixed by a simple sequence of poles and zeros.\footnote{The location of the poles 
and zeros does not fix the overall constant prefactor, but this can be fixed from 
knowledge of the geodesic limit computed in section \ref{sec:Geo}.}  
We do not have a simple derivation of the poles, but they are natural in the sense that they 
occur for values where  double-trace operators have the same dimension and spin 
as lowest-twist $T^n$s, and are indistinguishable from them in a conformal block decomposition.  
We will see that these poles arise as a common factor in all the OPE coefficients of 
leading twist $T^n$ operators, and therefore will be easier to obtain in our black-hole 
calculation on the first sheet without having to do any resummation of operators.  
By contrast, extracting the zeros at $\Delta=-m$ for $m=0, 1, \dots, 2(n-1)$ from a first 
sheet calculation requires resumming the infinite series of the lowest-twist 
operators, and therefore is more difficult.  

Suggestively, the zeros occur exactly at values of $\Delta$ such that the primary 
operator $\CO_L$ has degenerate descendants under the global conformal algebra.  
In \cite{Chen:2016cms}, it was shown how arguments based on null descendants 
can be used to completely fix the analogue of equation (\ref{eq:LTfromSW}) in $d=2$.  
Here we make some speculative comments about such an argument might work in  $d>2$.

In $d=2$, it is well-known  that at certain values of $\Delta$, the primary operator has null 
descendants that require its correlators to obey certain differential equations. 
These equations restrict the allowed operators that can appear in the $\CO_L \times \CO_L$ 
OPE, and constrain the position-dependence of correlators of $\CO_L$.  
To identify the necessary values of $\Delta$ and the corresponding differential equations, one 
has to look at the full Gram matrix of inner products of descendants of $\Delta$ at some level $m$ and 
 make sure that not only does the descendant in question have vanishing norm but 
moreover its overlap with all other descendants at that level vanish.  
Crucially, the resulting differential equations are satisfied on all sheets of the correlators, not 
only on the first sheet, so one can impose that only certain operators are allowed in the 
$\CO_L \times \CO_L$ OPE directly on the second sheet.  The conformal weights of allowed 
operators  in the OPE can be written explicitly at any value of $C_T$, but we only need them 
at infinite $C_T$ where they are especially simple and are controlled by the global conformal group.  
For 
$h\equiv\Delta_L|_{d=2}=-\frac{n}{2}$,  
the allowed intermediate operators have weights $h_I=0,-1 \dots, -n$.  
This is easy to see from the fact that at $C_T = \infty$, one of the degeneracy conditions 
is that the four-point function  must satisfy 
\be
d=2:~~~~~\partial_z^{n+1} \< \CO_L(0) \CO_L(z) \CO_H(1) \CO_H(\infty)\>=0  \ .
\label{eq:2dNullEq}
\ee
Since the leading small $z$ behavior of a block with weight $h_I$ is $\sim z^{h_I-2h}$, one 
immediately sees that (\ref{eq:2dNullEq}) can be satisfied only for the values of $h_I$ stated above.  
Therefore, a singularity on the second sheet with $z^{-m-1}$ or greater is forbidden when\footnote{We have multiplied by the prefactor $(z \bar {z})^{\Delta_L} = (z \bar {z})^{-m}$.}  
\be
d=2:~~~~~h =  0, -\frac{1}{2}, -1, -\frac{3}{2}, \dots, -\frac{m}{2} \ ,
\ee and thus its coefficient must vanish for 
all these values of $h$.  Conveniently, we did not need to use the form of the finite $C_T$ 
degeneracy conditions for this argument.

In $d>2$ (here $d=4$), let us assume that, in certain effective  limits,  
 one  can naturally expand the contribution of $T^n$ operators 
on the second sheet in some kind of ``generalized" primary operators, which may be called ``higher-$d$ Virasoro".    
 At infinite $C_T$, such generalized conformal blocks would reduce to $d=4$ global 
conformal blocks, which behave on the second sheet at small $z,\bar{z}$ like 
\be
d=4:~~~~~ \propto~ \frac{1}{(z \bar{z})^{\Delta}} \frac{z^a \bar{z}^b}{z-\bar{z}}
\ee
for values of $a,b$ that depend on the dimension and spin of the exchanged operator.
For $\Delta=-n$, the blocks again are annihilated by $\partial_z^{n+1}$.  Taking the small $z$ limit 
of an individual block, which is necessary to separate out the contributions from different blocks, and 
imposing $\partial_z^{n+1}=0$, we see that the allowed values of $a$ are $a=0, -1, \dots, -n$.  
To connect to our leading-twist limit, we  consider the blocks in the limit $\bar{z} \ll z$, where 
the power of $z$ is $z^{a-1}$.  
Therefore, the singularities $z^{-2m}$ must have  zeros at 
\be
d=4:~~~~~\Delta = 0, -1, \dots, -2(m-1) \ ,
\ee 
which is exactly what we see in our explicit computation!

These observations are encouraging, but  
 a potential problem  
could be that, in the first limit, $z \ll \bar{z}$, there 
is no obvious reason why there could {\it not} be a cancellation between the leading-twist 
and subleading-twist operators, yet we are attempting to constrain the contribution 
from the leading-twist operators alone.   
Nevertheless, it might be possible that the leading twists can be thought 
of as a special kind of conformal block.

\subsection{Factorization 
 Assumption
and 
Light-Light Limit}
\label{secFac}

So far, we have been working in the heavy-light limit, where the heavy 
operators $\CO_H$ in the 4-point function have dimensions $\Delta_H$ that 
scale like the central charge of the theory, whereas the light operators 
are parametrically smaller than the central charge.  
As a result, although our expression (\ref{eq:LTfromSW}) has complicated 
dependence on $\Delta_L$, all the dependence on $\Delta_H$ comes in simple 
powers together with $G_N$ in the combination $\Delta_H G_N \sim \frac{\Delta_H}{C_T}\sim f_0$. If $\Delta_H$ 
is taken to scale to infinity less quickly than $C_T$, then the true expression will have 
subleading corrections suppressed by powers of $\frac{1}{\Delta_H}$.  
 However,  we can try to infer these corrections from the large $\Delta_H$ limit that we already know.  The intuition 
is that the leading singularities on the second sheet can be thought of as coming from a single operator, and therefore their coefficients should factor into a product of a function of $\Delta_L$ only and a function of $\Delta_H$ only.    
In $d=2$, the second sheet OPE of null operators is simple and under rigorous control, and it is 
easy to see in that case that this ``single operator'' assumption is valid.  In $d>2$, the results of  \cite{Kravchuk:2018htv} suggest that the leading singularity  (\ref{eq:LTfromSW}) should be thought of as coming from a single ``light-ray'' operator,\footnote{In particular, as we will discuss in section \ref{sec:ReggePole}, the leading $T^2$ singularity is due to an isolated pole at $J=3$ in the analytic continuation of the lowest-twist $T^2$ OPE data.} so that factorization should hold here as well.

\begin{figure}[t!]
\begin{center}
\includegraphics[width=0.45\textwidth]{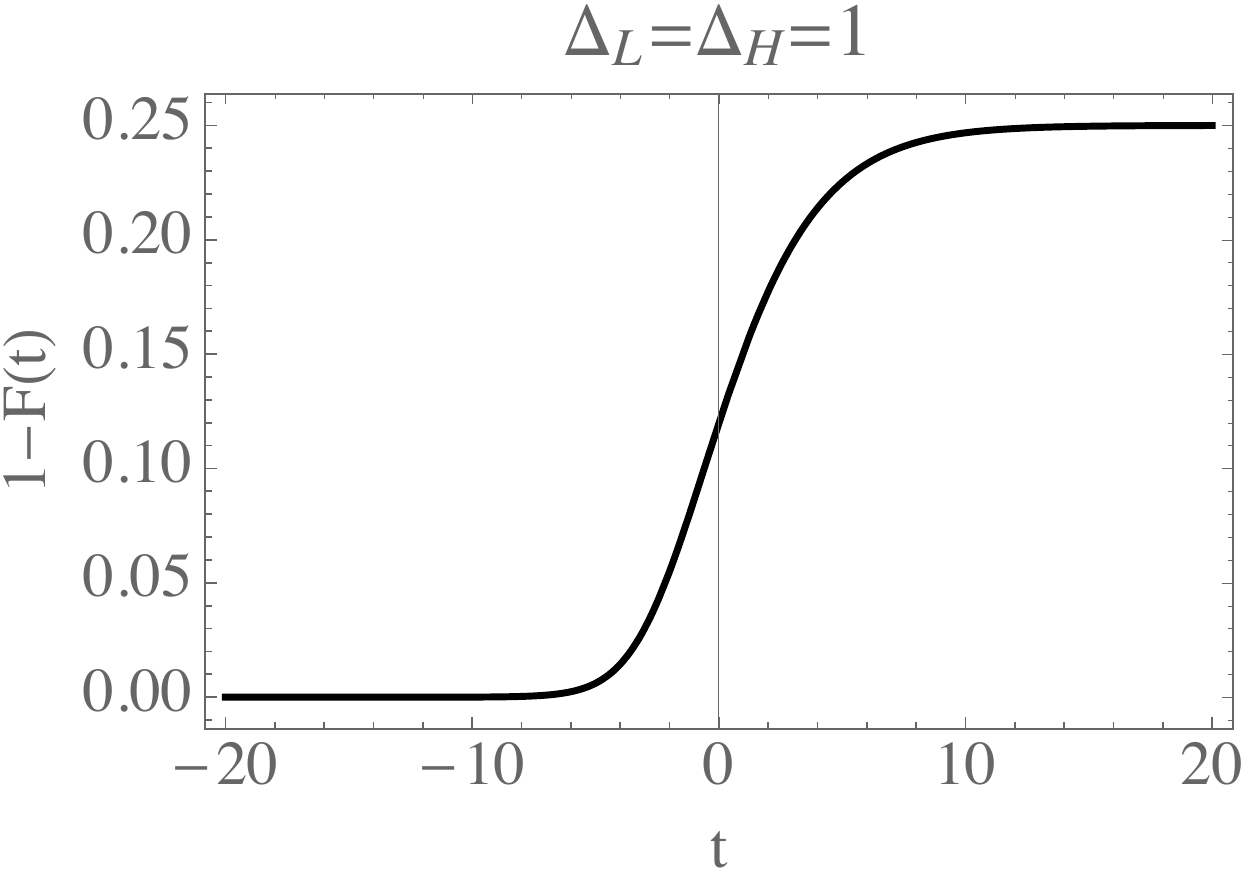}
\caption{$F(t) \equiv \frac{3}{4}+ \sqrt{\frac{x}{8\pi}} e^{x} K_0\left(x\right)|_{x=e^{-t}}$. }
\label{fig:Borel}
\end{center}
\end{figure}

Let us assume the coefficients do factorize. 
 One can then obtain the 
exact expression (at each order in a $1/C_T$ expansion but with $\Delta_H, \Delta_L  \sim \CO(1)$) 
by factorization together with the fact that the result must be symmetric in $\Delta_H \leftrightarrow \Delta_L$.  
Therefore, the full $\Delta_H$ dependence is identical to the $\Delta_L$ dependence we have already 
computed, and the resulting ``light-light'' correlator in the Regge limit at $\bar{z} \ll z$  in $d=4$ is 
\begin{equation}
\frac{G_{\rm Eik}(z,\bar{z})|_{h^n}}{G_{\rm Eik}(z,\bar{z})|_{h^0}} 
= (160 i \pi )^n 
\left( \frac{\bar{z}}{z^2 C_T} \right)^n \frac{  \Gamma \left(\Delta _L-n\right) \Gamma \left(\Delta _L+2 n-1\right)}{n! \Gamma \left(\Delta _L-1\right) \Gamma \left(\Delta
   _L\right)}\frac{  \Gamma \left(\Delta _H-n\right) \Gamma \left(\Delta _H+2 n-1\right)}{ \Gamma \left(\Delta _H-1\right) \Gamma \left(\Delta
   _H\right)} .
   \end{equation}
   We have used the fact that (\ref{eq:LTfromSW}) must be reproduced in the limit $\Delta_H \rightarrow \infty$. 
   
We would like to sum the above expression on $n$ in order to obtain a formula for the 
Regge limit at large $C_T$ with 
\be
\frac{\bar{z}}{z^2 C_T}~~~~ \rm {fixed} \ .
\ee
The naive sum has zero radius of convergence, but its Borel transform can be written in closed form, allowing 
us to obtain the following expression that interpolates between the region of large and small $\frac{\bar{z}}{z^2 C_T}$:
\be
\frac{G_{\rm Eik}(z,\bar{z})|_{h^n}}{G_{\rm Eik}(z,\bar{z})|_{h^0}} = \int_0^\infty dt e^{-t} \, _4F_3\left(\frac{\Delta _H}{2}-\frac{1}{2},\frac{\Delta _H}{2},\frac{\Delta _L}{2}-\frac{1}{2},\frac{\Delta
   _L}{2};1,1-\Delta _H,1-\Delta _L; \frac{2560 i \pi t \bar{z}}{C_T z^2} \right).
   \ee
The above formula simplifies in the limit $\Delta_L = \Delta_H=1$ to be\footnote{This limit is taken after fixing $n$ and before performing the resummation.}
\be
\frac{G_{\rm Eik}(z,\bar{z})|_{h^n}}{G_{\rm Eik}(z,\bar{z})|_{h^0}} 
=\frac{3}{4}+ \sqrt{\frac{x}{8\pi}} e^{x} K_0\left(x\right)
  , \qquad x \equiv  \frac{i C_T z^2}{5120  \pi \bar{z}}, 
\ee
where  $K_0$ is a Bessel function. The correlator is shown in this limit in Fig. \ref{fig:Borel}, as a function of $t= -\log(x)$.  In $d=2$, the coordinate transformation $z = e^{2 i \pi t/\beta}$ maps the theory to finite temperature, and the analogous plot to Fig. \ref{fig:Borel} shows the onset of ``chaos'' through the growth of out-of-time correlators at finite temperature \cite{Chen:2016cms}.  In $d=4$, the relation between zero and nonzero temperature is not just a coordinate transformation, and we do not have a corresponding simple physical interpretation of the behavior in  Fig. \ref{fig:Borel}.

\subsection{Geodesic and Exponentiation of ANEC}
\label{sec:Geo}

Let us here take the large $\Delta_{L}$ limit after taking other limits, so that the light 
operator travels along a geodesic. This limit 
automatically includes only the multi-stress-tensor contributions and 
 we no longer need to eliminate the contribution from the
double-traces made from the external operators. 

To see this from \eqref{Joao}, we take the leading
order in the saddle point approximation. Denote 
\begin{equation}
k\left(x,z,\bar{z}\right)\equiv2\mathbf{q}\cdot\mathbf{x}+h(\mathbf{p}\cdot\mathbf{x}) \ .
\end{equation}
The saddle point of the integral is given by $\partial_{x}k=0$. Plugging
in the solution $x_{*}$, we find 
\be
&&k\left(x_{*},z>\bar{z}\right)=-\sqrt{z\bar{z}}\left(1-\frac{64A\bar{z}}{(z-\bar{z})z}+\mathcal{O}\left(A^{2}\right)\right)\ , \\
&&k\left(x_{*},z<\bar{z}\right)= k\left(x_{*},z>\bar{z}\right)\big| z \leftrightarrow \bar z \ .
\ee
Kinematically, the difference between $z>\bar{z}$ and $z<\bar{z}$
maps to the fact that the probe geodesic can pass the source geodesic
from the left or the right side. 

Let us check that our result agrees with the single stress-tensor conformal block at linear order.
The single stress tensor block is
\be
g_{T}^{4d}=\frac{30}{z\bar{z}\left(z-\bar{z}\right)}\Big((-(z-6)z-6)\bar{z}^{2}\ln(1-z)
+z\left(6\left(z-\bar{z}\right)\bar{z}+z\left(\left(\bar{z}-6\right)\bar{z}+6\right)\ln\left(1-\bar{z}\right)\right)\Big).
\ee
On the second sheet, 
\begin{equation}
\lim_{z\rightarrow0,\ \frac{z}{\bar{z}}\ {\rm fixed}}\left.g_{T}^{4d}\right|_{(1-z)\rightarrow(1-z)e^{-2\pi i}}=-\frac{360i\pi\bar{z}}{\left(\bar{z}-z\right)z} \ ,
\end{equation}
\begin{equation}
\lim_{z\rightarrow0,\ \frac{z}{\bar{z}}\ {\rm fixed}}\left.g_{T}^{4d}\right|_{(1-\bar{z})\rightarrow(1-\bar{z})e^{-2\pi i}}=-\frac{360i\pi z}{\left(z-\bar{z}\right)\bar{z}} \ .
\end{equation}
The $z$, $\bar{z}$ dependence that appears at linear order in $A$
indeed agrees with the Regge limit of the stress-tensor block on the
second sheet. In particular, the two kinematic configurations of the
scattering correspond to continuing the 4-point function in $z$ or $\bar{z}$.

We now 
 replace 
the integral on the
shockwave surface by the saddle point value of its integrand. We focus on the $z>\bar{z}$ case; the result for $z<\bar{z}$ can
be obtained by $z\leftrightarrow\bar{z}$. 
We have 
\begin{equation}
G_{{\rm Eik}}~\propto~ \frac{1}{\left(1-\frac{64A\bar{z}}{(z-\bar{z})z}+\mathcal{O}\left(A^{2}\right)\right)^{2\Delta_{L}}} \ .
\end{equation}
Now take the geodesic limit with  
\begin{equation}
\Delta_{L}, ~ \Delta_{H},  ~ C_T \rightarrow\infty, ~~~ z,~ \bar{z}\rightarrow0,~~ 
{\rm with} ~~~ \frac{\bar{z}}{z},~ \frac{\Delta_{H}\Delta_{L}}{C_T~z}~~{\rm fixed} \ .
\end{equation}
Then 
\begin{equation}
G_{\rm Eik}\propto e^{-\frac{128A\bar{z}}{(z-\bar{z})z}\Delta_{L}}+\mathcal{O}\left(\frac{1}{\Delta_{L}}\right) \ .
\end{equation}
As the piece that is exponentiated arise from the single stress-tensor
block, which becomes the ANEC on the second sheet, we may write an operator form   
\begin{equation}
\ln {G}_{\rm Eik}~ \propto ~ \int dx^{-}T_{--} \ .
\end{equation}
We have therefore shown that 1) there exists a subset of OPE
data in the multi-stress-tensor sector that exponetiates the ANEC
on the second sheet; 2) the exponentiation of ANEC gives the leading
contribution in the geodesic Regge limit.

\section{Spherical Black Holes and  Lowest-Twist}
\label{sec4}

In this section, we consider solving the scalar field equation in a spherical black hole background  
with a general higher-order derivatives gravity action. 
After describing the field equation and the change of variables, we 
discuss the general structure of the perturbative solution in a near-boundary expansion. 
An all-orders proof of the universal lowest-twist operators involving two stress tensors will be given. 
We will derive recursion relations which lead to a closed-form 
prediction for the corresponding lowest-twist OPE coefficients.
Finally, we resum and perform an analytic continuation to the 2nd sheet 
to discuss the relationships with the shockwaves.  In this section, we will 
use $\Delta$ without a subscript to denote $\Delta_L$, for compactness.

\subsection{Field Equation}

Consider the bulk Euclidean action 
 \be
&& S_{\rm{tot}} =\int d^{4}x \sqrt{g} ~\Big( {\cal L}_\phi+ {\cal L}_{\rm grav} \Big)   \ ,  \\
&&{\cal L}_\phi = \frac{1}{2}(\partial \phi)^2 + \frac{1}{2} m^2 \phi^2  \ ,  ~~~ {\cal L}_{\rm grav}=  R+\Lambda+\sum_i  \alpha_i {\cal O} (R^2)+\sum_j \beta_j {\cal O} (R^3) 
+\sum_k \gamma_k{\cal O} (R^4)+\cdots     \nn
\ee 
where ${\cal L}_{\rm grav}$ is the most general higher-derivative gravity theory.  
${\cal O} (R^{\#})$ with powers $\#$ fixed denote all possible Lorentz invariants constructed out 
of Riemann tensor and metric;  indices $i,j,k$ represent independent invariants.
We focus on a spherical black hole with metric 
\be
\label{SHmetric}
ds^2= \big(1+r^2 f(r)\big) dt^2  + {dr^2 \over 1+ r^2 h(r) } + r^2 \sum_{i=1}^{3}d\Omega_i^2\ ,
\ee 
where $d\Omega_i^2$ with coordinates $(\theta_1, \theta_2 , \theta_{3})$ represents a unit 3-sphere. 
Using the rotation symmetry, we will remove $\theta_2$, $\theta_{3}$ dependence in the scalar and rename $\theta_1=\theta$ in the following. 
The asymptotic AdS boundary conditions imply \cite{Henneaux1985, Hollands:2005wt}
\be
f(r) = \frac{1}{\ell^2} - \frac{f_0}{r^4} + \dots, ~~~~~ h(r) = \frac{1}{\ell^2} - \frac{h_0}{r^4} + \dots \ .
\ee
We will set the AdS radius, $\ell$, to 1. 
In general, the higher-order corrections in $1/r$ depend sensitively on the higher-derivative terms in the gravity action.    
Conformal invariance in the boundary limit requires $h_0=f_0$ \cite{Fitzpatrick:2019zqz}.
The factor $f_0$ is give by 
\be
f_0 = {160 \over 3} \frac{\Delta_H}{C_T} \ ,
\ee 
at large $\Delta_H$ and $C_T$  but with the ratio fixed.  
A priori, the higher-order corrections to the functions $f(r)$ and $h(r)$ are generally independent. 
These higher-order terms, however, represent non-universal contributions. 
Since our final result will manifestly depend only on the coefficient $f_0$ of $r^{-4}$, and since $h_0=f_0$, we will 
simply set $f(r)=h(r)$ in what follows to keep the expressions simpler.\footnote{ 
We shall take the next-order term in $f(r)$ to be $f_4\over r^8$ as there is 
no non-trivial solution consistent with the conformal block decomposition 
if $f(r)$ or $h(r)$ has $1\over r^5$, $1\over r^6$, or  $1\over r^7$ structure \cite{Fitzpatrick:2019zqz}.}

To compute the two-point function of the light operator, we solve for the bulk-to-boundary 
propagator in the metric \eqref{SHmetric}. The bulk field equation is
\be
\label{eq:BulkEOMGen} 
\left( - \nabla^2 + m^2 \right) \Phi = 0 \ , ~~~~ m^2= \Delta (\Delta-d) \ , 
\ee
with the identification
\be
\Phi(r, x_1, x_2) \equiv \< \CO_L(x_1) \phi_L(r,x_2)\>_{\rm BH}  \ .
\ee   That is, we are interested in a two-point function of the light operator 
in a black hole geometry created by the heavy operator. 
We will mainly focus on the two-point function in the OPE limit, where two operators are close and the 
correlator depends only on the near-boundary behavior of the bulk fields and thus justifies a large $r$ expansion.   
As the OPE is a short distance expansion, i.e., an expansion in small $z$ and $z$ ̄, it can be taken regardless of the size of $f_0$. 
The expansion is therefore the conformal block decomposition of the 
four-point function, which is unambiguous, and this is how we will 
read off the OPE coefficients. 

In  \cite{Fitzpatrick:2019zqz}, two conjectures were made related to the spherical black hole:

{\bf --} {\it Weak conjecture: 
For each spin, the lowest-twist multi-stress-tensor  
operator for that spin is universal.}

{\bf --} {\it Strong conjecture:  
For each number $n$ of stress tensors, all the multi-stress-tensor  
operators with the lowest twist for that $n$ are universal.
}  

In this section, we will prove the weak version, where   the non-trivial cases with $J> 2$ 
have the lowest-twist operators with $two$ stress tensors and $J-4$ derivatives with suitable anti-symmetrization. 
We expect that a direct generalization on the present work will arrive at a 
general proof of the strong conjecture, but we will focus on two stress tensors here. 

To analyze a partial differential equation (PDE) such as the scalar field equation 
\eqref{eq:BulkEOMGen} in a black hole background, an 
important initial step often is to identify a suitable change of variables.
Let us first introduce
\be
A=(\sinh t)^2 \ , ~~~ B= (\sin \theta)^2 \ .
\ee 
The scalar field equation  can be written as 
\be
\Big(C_1 \del^2_r+ C_2 \del^2_A+C_3 \del^2_B + C_4 \del_r+ C_5 \del_A+C_6 \del_B - \Delta (\Delta-4)\Big) \Phi(r,A,B) =0 \ ,
\ee
where the coefficients are
\be
&& C_1= 1+r^2 f \ , ~~~~ C_2= {4A (1+A)\over 1+r^2 f} \ , ~~~~ C_3={4 B(1-B)\over r^2}\ , \\
&& C_4= {3\over r} + 5 r f + r^2 f' \ , ~~~~ C_5={2+4 A\over 1 + r^2 f}\ , ~~~~ C_6= {6 - 8 B \over r^2} \ .
\ee 
The pure AdS solution is 
\be
\label{pureads}
\Phi_{\rm AdS}=\Big({1 \over  2 \sqrt {(1 + r^2) (1 + A)} - 2 r \sqrt {1 - B} }   \Big)^\Delta \ .
\ee
We next consider the following change of variables:
\be
(r, t, \theta) \to  (r, A, B) \to (r, w, y) \ ,
\ee
where 
\be
w =  r \Big(2\sqrt{(1 + r^2) (1 + A)} - 2 r \sqrt{1 - B} \Big) \ , ~~~~y = 2 r^2 \Big(1 - \sqrt{1 - B}\Big) \ .
\ee   
The solution \eqref{pureads} has a simple form
\be
\Phi_{\rm AdS}=\big({r \over w}\big)^\Delta\ .
\ee
Although the field equation written in new variables $(w,y)$ becomes cumbersome and we shall not list it explicitly here, 
these new variables better organize the structure of the perturbative solutions discussed below. 
In particular, these new variables will help us isolate the lowest-twist contributions.

\subsection{Universal Lowest-Twist  of $T^2$}
\label{sec:UniversalLowestTwist}

In general, we can write the scalar solution as 
\be
 \Phi = \Phi_{\rm {AdS}}  G(r,  w , y)\ ,~~~~~~ G(r, w , y)= 1+G^{\rm{T}}(r, w, y)+ G^{\rm{\phi}}(r, w, y) \ , 
\ee  where the stress-tensor contributions, $G^{\rm{T}}$, is our main focus.   
The double-traces, $G^{\rm{\phi}}$, on the other hand, require certain 
interior boundary condition and thus we mostly ignore them in the following.  
As observed in \cite{Fitzpatrick:2019zqz}, the multi-stress-tensor conformal blocks 
are insensitive to the horizon boundary condition with either a planar black hole or  a spherical black hole. 
Our goal  is to show the OPE coefficients of the lowest-twist 
operators with two stress tensors are universal, i.e. they depend on $f(r)$ only through $f_0$, in the spherical black hole case.
The UV boundary conditions we impose are 
1) the standard $\delta$-function boundary condition; 2) the regularity at $w=1$.
The resulting stress-tensor contributions admit polynomial forms:
\be
\label{GTsph}
&&G^{\rm{T}}(r, w, y)={1\over r^4}\sum_{i=0,2,4,6,...}{G_i^{\rm{T}}(w, y)\over r^i} \ , 
\ee 
with, for instance, 
\be
G^T_{0}= \sum_{i=-1}^{2} a^{(0)}_i w^i+\sum_{j=-1}^1 b^{(0)}_j w^j  {y}\ ,~~~ 
G^T_{2}= \sum_{i=-1}^{3} a^{(2)}_i w^i+\sum_{j=-1}^{2} b^{(2)}_j  w^j  y+\sum_{k=-1}^{1} c^{(2)}_k w^k  y^2\ ,   
\ee   where $a, b, c$ are constant coefficients. 
It is straightforward to obtain similar polynomial forms at higher orders.\footnote{The variables 
adopted here are different from that in \cite{Fitzpatrick:2019zqz}, but the polynomial structure of the solution is formally the same. }
 
The proof of the universal lowest-twist OPE coefficients in 
the planar black-hole case \cite{Fitzpatrick:2019zqz} relies on a certain “decoupling” limit, 
imposed on the scalar field equation (in suitable variables), which 
leads a bulk reduced field equation that isolates the lowest-twist operators.  
We shall adopt a similar strategy with a spherical black hole. 
Before deriving the corresponding reduced field equation, let us first discuss some general structure. 

Using the variables $(r, w, y)$ defined above and focusing on stress tensors, we find 
that the perturbative solution can be recast into the following form:
\be 
\label{genphi}
\Phi= \Phi_{\rm{AdS}}\Big(1+ {q_2  w^2 y^2 + {\cal O}(y) \over r^8}+{q_3 w^2 y^3 
+ {\cal O}(y^2) \over r^{10}}+{q_4 w^2 y^4 + {\cal O}(y^3) \over r^{12}}+ \dots+ {\cal O}(f_0) \Big) \ . 
\ee 
The ${\cal O}(f_0)$ piece represents single stress-tensor contribution, which includes 
the ${\cal O}({1\over r^4})$ and ${\cal O}({1\over r^6})$ terms and also the $leading$-$y$ term at each order in $1\over r$. 
(For instance, $G^T_4\sim {\cal O}({1\over r^8})$ has the  leading-$y$ 
contribution $\sim y^3$, which is absorbed into ${\cal O}(f_0)$ part above.) 
In other words, in the spherical black hole case, the single stress-tensor 
contribution ``contaminates" the leading-$y$ part of the solution at each order in $1\over r$. 
As we are interested in the two stress-tensor contributions 
$\sim f^2_0$, we will not focus on the ${\cal O}(f_0)$ piece in the following.

Without explicitly solving the field equation, we can first consider the boundary 
limit of \eqref{genphi} and perform the conformal block decomposition to find 
relations between $q_2, q_3, q_4,...$ etc  and the OPE coefficients. By the 
boundary limit, we mean $r \to \infty$ with $t,\theta$ fixed and identify 
\be
\label{eq:FromGToPhi}
G(z,\bar z)=\<\CO_L(1) \CO_L(1-z)\>_{\rm BH} = \lim_{r\rightarrow \infty} r^{\Delta} \Phi(r, t,\theta)  \ .    
\ee 
Recall the map between coordinates 
\be
(1-z) = e^{t+i \theta} \ , ~~~ (1-\bar{z}) = e^{t-i \theta}\ ,
\ee
and the scalar 4-point function \cite{Dolan:2000ut}
\be
\label{CB}
&&
\langle {\cal O}_H(0) {\cal O}_L(z,\bar z) {\cal O}_L(1) {\cal O}_H (\infty) \rangle
= \sum_{\Delta_T, J}  c_{\rm OPE} (\Delta_T, J) g_{\Delta_T,J}(z,\bar{z}) .
\ee   
By substituting (\ref{genphi}) into (\ref{eq:FromGToPhi}) 
and expanding in terms of the blocks $g_{\Delta_T, J}$, one can verify that the coefficients $q_i$ 
determine the lowest-twist OPE coefficients associated 
with two stress tensors. Some explicit low-order relations are shown in Appendix \ref{App}. 
In the following, we will derive a reduced field equation 
whose solutions determine the coefficients $q_i$ to all orders. 
The reduced field equation will depend on $f(r)$ through $f_0$ only, which implies $q_i$ are protected.
Moreover, the map between $q_i$ and lowest-twist 
coefficients means that the lowest-twist coefficients are also universal.

To extract the lowest-twist coefficients, we have to look at the 
large $y$ limit of solution \eqref{genphi}, keeping subleading (in $y$) contributions.  
It is convenient to write
\be
\label{rwu}
{\Phi(r,w,u)\over \Phi_{\rm{AdS}}}=  1+ {1\over r^2} \Big(P(w,u)+ {Q(w,u)\over r^2} + \cdots \Big) \ , ~~~~~u= {y\over r^2} \ ,
\ee 
where $P(w,u)$ represents the leading-$y$ solution containing only 
one stress tensor; the function $Q(w,u)$ contains all the information 
of the lowest-twist operators with two stress tensors. 
Note that one can not simply set $P(w,u)=0$ by saying that one is only 
interested in two stress tensors as the subleading-order reduced 
field equation can have a term $\sim f_0 P$, which is of order $f^2_0$.  
At the field equation level, $P$ and $Q$ are generally entangled and we 
shall first obtain a leading-order reduced equation before going to the next order.

With $u$ fixed, plugging \eqref{rwu} into the full scalar 
field equation gives the following reduced bulk equation for $P$:  
\be
\Big(K_1 \del^2_w  +K_2 \del_u \del_w +K_3 \del^2_u+K_4 \del_w+K_5 \del_u  +K_6  \Big)P=- f_0   \Delta (\Delta+1) u (u-4),
\ee
and the next-order reduced bulk equation for $P$ and $Q$ reads
\be
&&\Big(M_1 \del^2_w  +M_2 \del_u \del_w +M_3 \del^2_u+M_4 \del_w+M_5 \del_u  +M_6  \Big)P \nn\\
&&+  \Big(N_1 \del^2_w  +N_2 \del_u \del_w +N_3 \del^2_u+N_4 \del_w+N_5 \del_u  +N_6  \Big)Q\nn\\
&&~~~ =  f_0 \Delta \Big(8 - 6 w + 3 u ( u + w-4) + 8\Delta + \big(3 (u-4) u + 4 (u-2) w + w^2\big) \Delta\Big).
\ee
The coefficients $K_i,M_i$ and $N_i$ are simple but we relegate them to Appendix \ref{App} to avoid clutter.

As multi-stress tensors are fixed by UV boundary conditions and the 
above reduced field equations depend on $f(r)$ through $f_0$ only, we arrive at an all-order proof 
of the lowest-twist universality with two stress tensors in the spherical black hole case. 
We emphasize that these computations go beyond the geodesic approximation. 

Given the reduced field equations and the general structure of the 
purturbative solution \eqref{rwu}, we next derive recursion relations. 
Write
\be
P(u,w)=a_{n,m} u^n w^m   \ , ~~~~~ Q(u,w)=b_{n,m} u^n w^m    \ .
\ee 
We find, after matching powers, 
\be
&&a_{n,m}= {(m - n-1) a_{n-1,m}+  2 (m-2) a_{n,m-1} \over 2 (m - 2 n-2)} \ , \\
&&b_{n.m}=
\Big[ f_0 \kappa \big(\kappa+1\big) \big(a_{n-2,m+1} - 4 a_{n-1,m+1}\big) \nn\\
&&~~~~~~~~~~ +2 \kappa  \big(m - n-3\big)b_{n-1,m} + 4 \big(\kappa-1\big)\big(m-3\big)  b_{n, m-1} \Big]\nn\\
&&~~~~~~~~~~~~~ \times {1\over 4 \kappa \big(m - 2 n-4 \big)} \ , ~~~~~~~~\kappa\equiv m - \Delta \ ,
\ee
with initial values 
\be
a_{1,1}= - {\Delta \over 30}  f_0\ , ~~ a_{1,0}= -  {\Delta\over 10}  f_0 \ , ~~ a_{1,-1}= - {\Delta\over 5}  f_0\ , ~~ a_{2, -1}= - {3\Delta \over 140}  f_0  \ ,
\ee 
and $a_{n,m}=0$ if $n<1$, $m<-1$ or $m>1$; $b_{n,m}=0$ if $n<0$, $m<-2$ or $m>2$.    
The recursion relations allow us to compute the OPE coefficients effectively;  coefficients up to $J=18$ are listed in Appendix \ref{App}.  

\subsection{Resummation and Continue to the 2nd Sheet}

A formula that agrees with all the lowest-twist OPE coefficients we have computed is
\be
\label{cformula}
 c_{T^2, J} &=& \gamma_J    \frac{ 9 (\Delta +2) (\Delta +1)+\Delta  (\Delta +1) x_J+\Delta  (\Delta +2) y_J}{\Delta -2} \Delta  f_0^2\ , 
\ee
where
\be
 \gamma_J&=&\frac{J(J-2)  \Gamma (J-3) \Gamma (J+1)}{(J+2) (J+4) (J+6) \Gamma (2 J+2)}\ , \\
 x_J &=& \frac{3}{16} (J-1) (J-3) (J+4) (J+6) \ , \\
 y_J &=& -\frac{1}{8} (J-2) (J-3) (J+5) (J+6) \ .
\ee

We next resum the contribution from the infinite set of such 
operators and analytically continue to the Regge limit.  In the 
lightcone limit $\bar{z} \ll 1 $, the conformal blocks take the form  \cite{Dolan:2000ut}
\be
(z \bar{z})^\Delta g_{\tau, J}(z,\bar z) = (z \bar{z}) ^{\frac{\tau}{2}} z^J F(J + \frac{\tau}{2}, J+ \frac{\tau}{2}, 2J + \tau, z) \ .
\ee
The hypergeometric function has a branch cut from $z=1$ to $z=\infty$.  
Moving $z$ through this cut and then taking $z \sim 0$, and using the fact 
that $\tau=4$ for the minimum twist $T^2$ operators we are considering,  
we have, at small $z$ on the second sheet,
\be
(z\bar{z})^\Delta g_{4,J} &\sim& 
  2\pi i\frac{\bar{z}^2}{z^{J+1}}  \frac{\Gamma(2J+4)\Gamma(2J+3)}{\Gamma^4(J+2)}\big( 1 + \CO(z) \big) .
\ee 
Putting this together with the OPE coefficients and keeping the most singular term (in $1/z$) for 
each conformal block, we find that the $T^2$ contribution to the four-point function is
\begin{equation}
\label{eq:GT2SS}
(z \bar{z})^\Delta G_{T^2} \stackrel{z \sim 0}{\sim}  - \left( \frac{\bar{z}}{z^2} \right)^2 \frac{9 \pi ^2 \Delta  f_0^2}{2 (\Delta -2)} \Big( (\Delta +1) (\Delta +2)   
+ \CO(z)
 \Big) \ . 
\end{equation}
 Note that the most singular term $z^{-2}$ is softer than the singularity $z^{-3}$ of 
even the first $T^2$ operator, $J=4$.  The leading term at small $z$ agrees exactly 
with (\ref{eq:LTfromSW}) at $n=2$, completing 
the check that the two methods agree in their region of overlap.

\subsection{$T^2$ Regge Pole}
\label{sec:ReggePole}

\begin{figure}[t!]
\begin{center}
\includegraphics[width=0.5\textwidth]{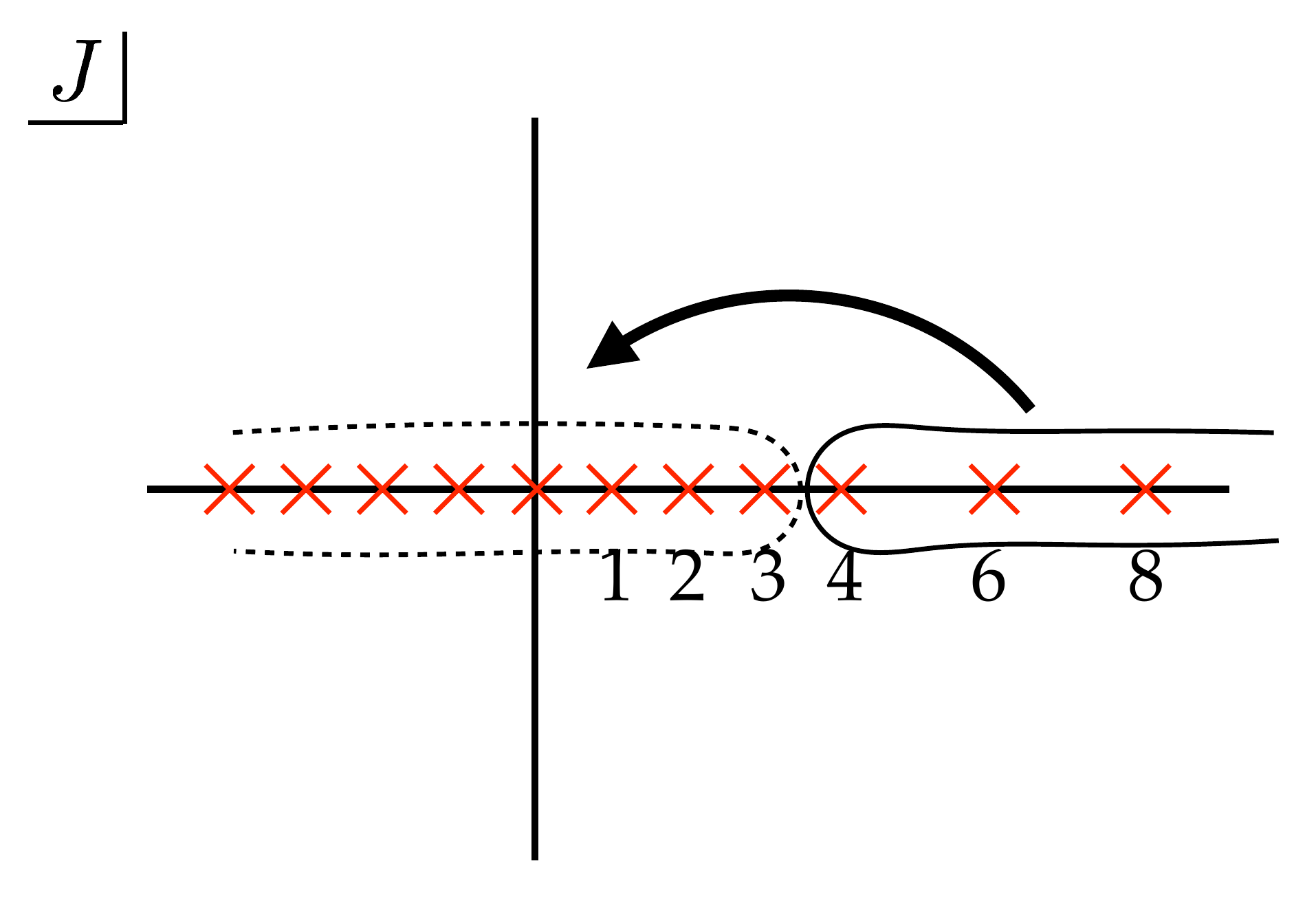}
\caption{Deforming the contour in the complex $J$-plane.}
\label{fig:ReggePoles}
\end{center}
\end{figure}

In the above derivation, we used the explicit knowledge of the lowest-twist $T^2$ OPE 
coefficients for all $J$. Was it necessary to know this entire functional form in order to 
deduce the leading singularity at $\bar{z} \ll z\ll 1$ in the Regge limit? If it were, then 
one might in principle be able to work backwards and deduce the function $c_{T^2,J}$  
OPE coefficients purely from the shockwave analysis.  However, this does not appear to 
be possible.  Instead, the part of the OPE coefficient that is fixed by the shockwave limit 
is the residue of the pole at $J=3$.\footnote{We thank S. Caron-Huot for pointing this out to us.}
  To see this explicitly, write the sum over the $T^2$ operators as an integral as follows:
\begin{equation}
G_{T^2}(z,\bar{z}) = \sum_{J=4}^\infty c_{T^2,J} g_{\tau,J}(z,\bar{z}) 
\sim \int \frac{dJ }{2\pi i} \frac{\pi}{2\sin \pi J} c_{T^2,J}\left[g_{\tau,J}(z,\bar{z}) + g_{\tau,J}\big(\frac{z}{z-1}, \frac{\bar{z}}{\bar{z}-1}\big)\right] \ , 
\end{equation}
where the integration contour is over the solid curve in Fig. \ref{fig:ReggePoles}.  
For integer spin, the second term in brackets is $(-1)^J$ times the first term and 
therefore cancels it for even spin, see e.g. \cite{Cornalba:2007fs,Costa:2012cb}.  
After analytic continuation in $z$, on the second sheet the blocks behave like
\be
(z\bar{z})^\Delta g_{\tau,J}(z,\bar{z}) \sim 2\pi i \frac{z^{2-J} }{z-\bar{z}} \left( \frac{\bar{z}}{z} \right)^{\frac{\tau}{2}} \frac{\Gamma(\tau+2J)\Gamma(\tau+2J-1)}{\Gamma^4(\frac{\tau}{2}+J)}\ , 
\ee
so that the dominant contributions at small $z,\bar{z}$ with $z/\bar{z}$ fixed are from 
the largest values of spin.  It is therefore advantageous to analytically continue the 
contour in $J$ to the left, as shown in Fig. \ref{fig:ReggePoles}.  The dominant 
contribution can then be read off from the pole with the largest value of $J$ 
contained within the contour, in this case $J=3$: 
\be
(i\pi)^2 \frac{\bar{z}^2}{z^{1+J}} \frac{\Gamma(4+2J)\Gamma(3+2J)}{\Gamma^4(2+J)} c_{T^2,J} \stackrel{J \sim 3} \sim - \frac{1}{J-3} \left( \frac{\bar{z}}{z^2} \right)^2 \frac{9 \pi ^2 f_0^2\Delta (\Delta +1) (\Delta +2) }{2 (\Delta -2)} ,
\ee
from the formula (\ref{cformula}). One can think of this pole and its residue as the ``intersection'' of the shockwave and black hole methods for the $T^2$ operators.  That is, it can be read off from the explicit form of $c_{T^2,J}$ from the block hole computation and then analytically continuing,\footnote{In order to deform the contour in the $J$ plane as described, the OPE coefficients must be analytically continued from real integer values to complex values in such a way that they decay sufficiently rapidly at infinity.  The 
form 
(\ref{eq:T2OPE}) 
has 
this behavior.} or it can be read off from the shockwave computation by inspection of the leading singularity in the $\bar{z} \ll z\ll 1$ regime of the Regge limit at $\CO(\frac{1}{C_T^2})$ from \eqref{eq:EikHn}. 
 The $O_H O_H$ and the $O_L O_L$ OPE each contributes an ANEC at order $T^2$ on the second sheet with $\bar{z} \ll z\ll 1$.  
These two ANEC operators contain a spin-3 operator \cite{Hofman:2008ar,Kologlu:2019mfz} and we find its contribution here.

\section{Discussion}

In this work, we  discuss the  connection between the universality in the 
lowest-twist limit (with a black hole) and  the universality in the Regge limit (with a shockwave), 
as distinct pieces of a connected universality region in $d>2$ CFTs.\footnote{While we focus on $d=4$, we do 
not expect any major change in other dimensions, but it is possible that the structure in odd dimensions 
is more involved compared to that in even dimensions.} Our results do not rely on  
unitarity or supersymmetry, and they are insensitive to the 
higher-curvature terms in the gravity action. 
Evidently, in the large space of holographic CFTs above 
two dimensions, there is a rather special  region where the CFT data are protected in 
the sense that they do not depend on  the additional model-dependent parameters  represented by such terms.

What is the ``boundary" of this space of universality?   
The analysis of the present work provides additional hints 
but a complete answer to this question is still lacking. 
In the planar black-hole case, it was known that the sub-leading twist 
OPE coefficients are not universal \cite{Fitzpatrick:2019zqz}. 
The structure of a spherical black hole is richer and a more detailed 
classification is still needed. In particular, we are able to prove the 
weak conjecture, which states that, for each spin, the lowest-twist 
multi-stress-tensor operator for that spin is universal. 
 It would be interesting to generalize our computations to prove the 
strong conjecture, which states that all the multi-stress-tensor operators 
with the lowest twist for $n$ stress tensors are universal. 
 It is possible that an even stronger version of universality exists in $d>2$ 
CFTs and so far we have only seen the tip of the iceberg. 
(See also the appendix A of \cite{Fitzpatrick:2019zqz} for related remarks.) 
 It will be interesting to carve out  more precisely the region where the universality holds.\footnote{We have adopted
  heavy operators and a large central charge. It would be very interesting to study the universalities away from these limits.}

For instance, in both shockwave and black-hole 
computations,  matter fields in the bulk are not included.  
To our knowledge, the shockwave solution in the presence 
of bulk matter fields has not been considered in the literature. 
Heavy matter fields can be integrated out and their effects can be absorbed into coefficients of irrelevant operators, so we would expect any corrections to be at least nonperturbatively suppressed in a limit that the masses of all such fields becomes large. It would be useful to explicitly compute their effects as a function of mass and test this expectation, and understand in detail the behavior of such corrections in both the shockwave and black hole regimes.\footnote{See  \cite{Li:2019tpf} for recent work in this direction.}

We would also like to understand from the CFT perspective {\it why} 
these universalities exist at all.  In $d = 2$, the Virasoro algebra essentially 
determines the related structures. Above two dimensions, the 
stress tensors generally do not form a universal algebra.  
However, the results in $d>2$ in the lowest-twist limit and also in the 
Regge limit share striking similarities with $d=2$ CFTs. 
 It is natural to ask, in the region where these universalities hold, if 
one could give a CFT derivation.  
A recent work \cite{Huang:2019fog} has shown that a Virasoro-like stress-tensor 
commutator structure effectively emerges near the lightcone in $d>2$ CFTs. 
It would be interesting 
to search for a connection and provide an algebraic approach to these universalities.
Relatedly, it will be also interesting to see if these universal results in $d>2$ CFTs 
can link to recent works on ANEC and lightray operators 
\cite{Casini:2017roe, Kravchuk:2018htv, Kologlu:2019mfz,Cordova:2018ygx, Belin:2019mnx, Kologlu:2019bco, Balakrishnan:2019gxl}.

Another potential way of trying to synthesize the 
``lowest-twist black hole'' and 
``shockwave'' limits of the $T^n$ confomal blocks is  
to focus on how they could be embedded within a larger structure that contain both of them as limits.  
One option for such a structure is the contribution of all $T^n$ operators in the limit where the bulk theory simply is Einstein gravity. 
 In the heavy-light limit, these contributions can equivalently be defined as the parts of the multi-stress-tensor partial wave that are proportional to powers of $f_0$. 
One may therefore define a special ``Einstein block'' that includes all $T^n$ operators  with  $f_n$s in the metric set to zero except $f_0$. By definition, it is insensitive to higher-curvature corrections.   
Nevertheless, we  emphasize that the spirit of our present paper has been to understand results which hold beyond Einstein gravity, and thus such a special conformal block appears somewhat ad hoc -- why should we set the $f_n$s with $n>0$ to vanish, when any other choice for their values would also define a larger structure that includes the lowest-twist black hole and shockwave $T^n$s as limits?  
 It would be nice to understand if a particular choice follows from purely CFT reasoning, without imposing a large gap in dimensions.

\subsection*{Acknowledgments}

We thank 
N. A.-Jeddi, B. Balthazar, S. Caron-Huot, M. Cho, S. Collier, E. Dyer, P. Gao, T. Hartman, J. Kaplan, E. Katz, S. Kundu, Z. Li, 
D. Meltzer, B. Mukhametzhanov, V. Rodriguez, E. Shaghoulian, D. S-Duffin, N. Su, A. Tajdini, M. Walters, J. Wu and X. Yin 
for discussions.
ALF and KWH were supported in part by the US Department of Energy Office of Science under Award Number DE-SC0015845 
and in part by the Simons Collaboration Grant on the Non-Perturbative Bootstrap, and ALF in part
by a Sloan Foundation fellowship. 
DL was supported by the Simons Collaboration Grant on the Non-Perturbative Bootstrap.
\\

\appendix 
\section{OPE 
and Bulk Coefficients}
\label{App}

\addtolength{\parskip}{0.75 ex}
\jot=0.75 ex

In this appendix, we provide explicit expressions for some technical details in section \ref{sec:UniversalLowestTwist}.

First, we list the lowest-twist coefficients with two stress tensors from $J=4$ up to $J=18$,  
\be
&& c_{\rm OPE}(8,4)= \frac{\Delta  (\Delta  (7 \Delta +6)+4) f_0^2}{201600 (\Delta -2)}\ , ~~~~~~~~~~~~~~~ c_{\rm OPE}(10,6)= \frac{\Delta  (\Delta  (33 \Delta -7)+4)  f_0^2}{38438400 (\Delta -2)} \ ,\nn \\
&& c_{\rm OPE}(12,8)= \frac{\Delta  \left(286 \Delta ^2-157 \Delta +12\right) f_0^2 }{8576568000 (\Delta -2)} \ ,~~~~~~ c_{\rm OPE}(14,10)= \frac{\Delta  \left(325 \Delta ^2-229 \Delta +6\right)  f_0^2}{219011240448 (\Delta -2)} \ , \nn\\
&& c_{\rm OPE}(16,12)= \frac{\Delta  \left(425 \Delta ^2-336 \Delta +4\right)  f_0^2}{5996736345600 (\Delta -2)} \ , ~~~~~~ c_{\rm OPE}(18,14)= \frac{\Delta  (\Delta  (2261 \Delta -1907)+12)  f_0^2}{638835994368000 (\Delta -2)}\ ,\nn\\
&& c_{\rm OPE}(20,16)= \frac{\Delta  (\Delta  (3724 \Delta -3271)+12)  f_0^2}{20422788194952000 (\Delta -2)}\ , ~~ c_{\rm OPE}(22,18)= \frac{\Delta  (\Delta  (483 \Delta -436)+1)  f_0^2}{50232688912560000 (\Delta -2)} \ . \nn
\ee 

The relation between the first several coefficients $q_i$ in 
section \ref{sec:UniversalLowestTwist} and the lowest-twist $T^2$ coefficients are
\be
&&q_2 = 16 c_{\rm OPE} (8,4) + {\cal O}(f_0)\ ,  \\
&&q_3 = {184\over 13} c_{\rm OPE} (8,4) - 64 c_{\rm OPE} (10,6)+ {\cal O}(f_0)\ , \nn \\
&&q_4 = {521\over 65} c_{\rm OPE} (8,4) - {1232\over 17} c_{\rm OPE} (10,6)+ 256 c_{\rm OPE} (12,8)+ {\cal O}(f_0)\ , \nn \\
&&q_5 = {4078\over 1105} c_{\rm OPE} (8,4) 
- {16164\over 323} c_{\rm OPE} (10,6) + 
 {7424\over 21} c_{\rm OPE} (12,8)- 1024 c_{\rm OPE} (14,10)+ {\cal O}(f_0)\ , \nn 
\ee 
and so on.   
The ${\cal O}(f_0)$ piece on the right-hand side of the relations represent 
the single stress-tensor contribution which is fixed by the Ward identity and depends only on 
$\Delta$ and $f_0$. More precisely, $c_{\rm OPE} (4,2)={\Delta\over 120} f_0$.

 Finally, the coefficients $K_i, M_i$ and $N_i$ from the bulk reduced field equations are
\be
&& K_1=2 w^3 (u + 2 w-2) \ , ~~~~~~ K_4= 2 w^2 \Big( (2-u) (\Delta+1) -2 w ( \Delta-1)\Big) \ , \\
&& K_2= 2 u w^2 (4-u)  \ ,~~~~~~~~~~~~ K_5= 2 u  w \Delta  (u-4) \ , \nn \\
&& K_3=0 \ ,~~~~~~~~~~~~~~~~~~~~~~~~~~~ K_6= 4 w \Big(w (\Delta-1) + ( u-2) \Delta\Big)  \ ,\nn 
\ee
and  
\be
&& M_1=w^2 \big( w^2 -   f_0 u(4-u) \big) \ , ~~~~~~~N_1=2 w^3 (u + 2 w-2) \ ,\\
&& M_2= 0 \ , ~~~~~~~~~~~~~~~~~~~~~~~~~~~~~~~~~ N_2=  2 (4 - u) u w^2 \ ,\nn\\
&& M_3=(4 - u) u w^2\ ,~~~~~~~~~~~~~~~~~~~~ N_3= 0 \ , \nn\\
&& M_4= -w \big(w^2    -   2 f_0 (4 - u) u \Delta \big) \ ,~~N_4= -2 w^2 \Big(u ( \Delta+3) + 2 ( w - \Delta + w \Delta-3)\Big)\ ,  \nn\\
&& M_5= 3 (2 - u) w^2\ , ~~~~~~~~~~~~~~~~~~~~N_5= 2 (u-4) u w \Delta \ , \nn\\
&& M_6=   -   f_0 (4 - u) u \Delta (\Delta+1) \ , ~~~~~~N_6= 8 w (u + w-2) \Delta  \ . \nn
\ee 


\bibliographystyle{utphys}
\bibliography{ShockandBHBib}

\end{document}